\documentclass[12pt]{spieman}  
\usepackage{amsmath,amsfonts,amssymb}
\usepackage{graphicx}
\usepackage{setspace}
\usepackage{tocloft}
\usepackage{lineno}
\usepackage{bm}
\makeatletter
\DeclareRobustCommand{\captioncite}[1]{%
  \begingroup
  \@fileswfalse
  \cite{#1}%
  \endgroup
}

\title{Time-Varying Photonics: A Review of Theory, Phenomena, and Experimental Platforms}

\author[a]{Zhaohui Dong}
\author[a,b]{Xianfeng Chen}
\author[a,*]{Luqi Yuan}
\affil[a]{State Key Laboratory of Photonics and Communications, School of Physics and Astronomy, Shanghai Jiao Tong University, Shanghai 200240, China}
\affil[b]{Collaborative Innovation Center of Light Manipulation and Applications, Shandong Normal University, Jinan 250358, China}

\cftpagenumbersoff{figure}
\cftpagenumbersoff{table} 
\begin{document} 
\maketitle

\begin{abstract}
Time-varying photonics opens new avenues for controlling waves through engineering material properties in time. 
In this review, we introduce the basic concepts and theoretical tools for describing temporal interfaces, time-periodic systems, and space-time media, as well as review representative experimental approaches across different photonic and wave platforms. 
We further discuss characteristic phenomena enabled by temporal modulation, including coherent wave control, parametric amplification, nonreciprocity, and topological effects. 
This review provides a unified pathway for understanding time-varying photonics and outlines its current challenges and future opportunities in both fundamental physics and practical applications.
\end{abstract}


{\noindent \footnotesize\textbf{*}Luqi Yuan,  \linkable{yuanluqi@sjtu.edu.cn} }

\begin{spacing}{2}   

\section{Introduction}
Artificial materials have greatly reshaped modern optics over the past decades, which establish a new paradigm for manipulating light fields through engineering material structure and enable precise control over the propagation, scattering, and emission of electromagnetic waves. 
Photonic crystals exploit spatial periodicity to tailor dispersion and photonic bandgaps, providing a powerful tool for molding the flow of light \cite{yablonovitch1987inhibited, john1987strong, ho1990existence, joannopoulos1997photonic, chan2025essay}. 
Metamaterials and metasurfaces extend this idea through subwavelength structuring, giving rise to extraordinary electromagnetic properties beyond those in natural materials \cite{pendry2000negative, smith2004metamaterials, yu2011light, ni2012broadband, sun2012gradient, kadic2013metamaterials, chen2016review, kadic20193d, wu2025intelligent}. 

While these developments have largely focused on engineering the spatial structure of materials, varying their properties in time offers a complementary approach to controlling electromagnetic waves. The study of time-varying electromagnetic systems has a long history, dating back to the middle of the twentieth century. Early developments in linear time-varying network theory and parametric circuits established fundamental approaches to analyzing and manipulating waves in dynamically varying systems \cite{zadeh1950frequency,cullen1958travelling,tien1958parametric}. Pioneering investigations of electromagnetic propagation in media with time-dependent constitutive parameters established the foundations of temporal scattering \cite{morgenthaler1958velocity,felsen1970wave,fante1971transmission}. In parallel, studies of traveling-wave and space-time-periodic modulations revealed the effects of temporal variation on wave coupling, dispersion, and stability \cite{oliner1961wave,cassedy1963dispersion,cassedy1967dispersion}. Other early studies explored parametric amplification, nonreciprocity, and wave propagation in nonstationary dispersive media \cite{holberg1966parametric,kamal1960parametric,stepanov1976dielectric,shvartsburg2005optics}. These pioneering works laid an important foundation for modern time-varying photonics \cite{ptitcyn2023tutorial}.
Building on these foundations, recent advances in material engineering and modulation technologies have enabled increasingly precise control over electromagnetic waves through temporal modulation \cite{engheta2023four,galiffi2022photonics,yin2022floquet,hayran2023using,yu2026fundamentals,luo2026temporally}. A natural question then arises: what new possibilities emerge when artificial structure is extended to the time domain?

Such an extension is not a simple replacement of space by time. Instead, the different symmetries of spatially and temporally varying systems lead to different conserved quantities and physical constraints, resulting in fundamentally distinct light-matter interactions. 
For temporal structures, the continuous time-translational symmetry is broken, allowing electromagnetic waves to exchange energy with external sources through modulation. 
This leads to a variety of novel phenomena, including but not limited to nonreciprocal light propagation \cite{sounas2017non, shaltout2015time, taravati2020full, wang2020nonreciprocity, chen2019nonreciprocal}, frequency conversion \cite{caspani2016enhanced, alam2016large, guo2019nonreciprocal, zhou2020broadband, khurgin2020adiabatic}, and extreme energy manipulation \cite{mirmoosa2019time, li2021temporal, hayran2024beyond, ciabattoni2025observation}. 
Temporal periodic modulation further gives rise to Floquet band structures \cite{oka2019floquet, rudner2020band}, photonic time crystals (PTCs) \cite{asgari2024theory, boltasseva2024photonic, dong2026photonic, guo2026plasmonic}, and synthetic frequency dimensions \cite{yuan2021synthetic, yu2025comprehensive}, while the combination of spatial and temporal modulation enables exotic phenomena such as spatiotemporal crystals \cite{smits2018observation, sharabi2022spatiotemporal}, luminal gratings \cite{pendry2021gain, pendry2021gain1, horsley2023quantum}, and space-time metasurfaces \cite{hadad2015space, zhang2018space, wang2020theory, sisler2024electrically}, which has no counterpart in static systems. 
Therefore, the inclusion of temporal modulation offers a new degree of freedom for controlling light, which is fundamentally distinct from conventional spatial structuring.

This review aims to provide a pedagogical introduction to the basic concepts, theoretical tools, and physical phenomena in time-varying photonics. 
We begin by formulating scattering at temporal interfaces and deriving the corresponding scattering matrices from interface conditions. We then discuss possible opportunities in temporal slabs comprising multiple interfaces and topological phenomena related to temporal interfaces. 
We then extend this picture from a single temporal discontinuity to periodically modulated systems, where Floquet theory and the transfer matrix method are employed as the basic theoretical tools. 
With these useful tools in hand, we can analyze how temporal modulation reshapes photonic band structures and wave dynamics, with PTCs and synthetic frequency dimensions serving as representative examples. 
We further move beyond purely temporal modulation to space-time media, where the interplay between spatial and temporal variations results in richer wave phenomena. 
Finally, we summarize the representative experimental platforms, highlighting how time-varying photonics can be implemented in various wave systems.

\section{Wave scattering at temporal interfaces}
\subsection{Basic physical picture}
\begin{figure}[tbp]
    \centering
    \includegraphics[width=8cm]{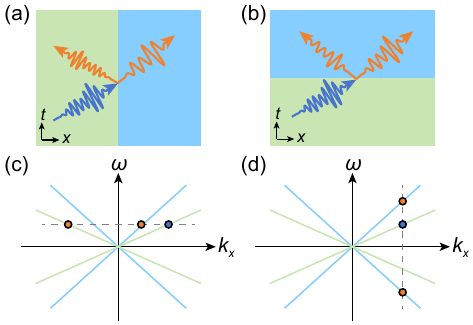}
    \caption{Scattering at (a) a spatial interface and (b) a temporal interface in a one-dimensional system}: when an incident wave encounters a spatial discontinuity, it generates a refracted wave and reflected wave propagating in different media, while a temporal discontinuity produces refracted wave and reflected wave propagating in the same medium. (c)-(d) Illustration of the scattering on dispersion diagrams, where the green and blue lines denote the dispersion of the two media. Yellow (orange) dots denote the incident (refracted and reflected) waves. Here, $k_x$ denotes the $x$-component of the wavevector, with positive and negative values corresponding to opposite propagation directions.\label{figure.1}
\end{figure}

We begin with wave scattering at a temporal interface, which provides a fundamental building block for understanding more complex time-varying electromagnetic media. A temporal interface is the time-domain counterpart of a spatial interface, occurring when the properties of a medium change abruptly in time rather than across a spatial boundary. Fig. \ref{figure.1} compares the scattering processes at spatial and temporal interfaces using space-time and dispersion diagrams. At a spatial interface, an incident wave generates reflected and transmitted waves that propagate on opposite sides of the boundary [Fig. \ref{figure.1}(a)]. The reflected wave propagates backward, with its spatial profile inverted relative to the incident wave. In contrast, a temporal interface involves a simultaneous change of the medium globally. The incident wave generates time-refracted and time-reflected components, both propagating in the new medium but in opposite spatial directions [Fig. \ref{figure.1}(b)]. Importantly, time reflection does not imply backward propagation in time. Instead, the time-reflected wave reverses its spatial propagation direction while retaining the original spatial profile at the temporal interface.

The dispersion diagrams provide a complementary interpretation of scattering in terms of the allowed wave states. 
For a spatial interface, the system breaks spatial translational symmetry while preserving time-translation symmetry, leading to the conservation of energy. 
Consequently, the incident wave projects horizontally, generating a spatial refraction with a positive $k_x$ and a spatial reflection with a negative $k_x$ [see Fig. \ref{figure.1}(c)]. 
For a temporal interface, the system undergoes a spatially uniform change in the time domain, hence preserving spatial translational symmetry while breaking time-translational symmetry. 
Consequently, a temporal interface holds momentum conservation and does not conserve energy. 
The incident wave therefore projects vertically along the frequency axis, as illustrated in Fig. \ref{figure.1}(d), resulting in a time-refracted wave and a time-reflected wave that carry opposite frequencies. 
Note that the negative frequency of a time-reflected wave shall not be regarded as negative energy. Instead, the time-reflected wave represents an opposite temporal phase evolution with respect to the time-refracted wave \cite{mendoncca2002time, moussa2023observation}. 
The discussion here points to the fundamental difference between a spatial reflection and a temporal reflection: although they are both backward-propagating, spatial reflection reverses the wavevector, while temporal reflection reverses the temporal phase evolution, which originates from the preserved and broken symmetries at the interface.

\subsection{Interface conditions and temporal scattering}

With the basic physical picture in mind, we can further analytically model the wave scattering at a temporal interface. 
We consider a temporal interface at $t=t_0$, where the the electromagnetic properties of the medium are abruptly changed. 
We start from the Maxwell's equations
\begin{equation}
    \begin{aligned}
    \nabla\times \bm{E} &= -\frac{\partial \bm{B}}{\partial t},\\
    \nabla\times \bm{H} &= \frac{\partial \bm{D}}{\partial t}+\bm{J},
    \end{aligned}\label{eq1}
\end{equation}
where $\bm{E}, {H}$ are the electric and magnetic fields, $\bm{B}, \bm{D}$ are the magnetic flux density and electric displacement, and $\bm{J}$ is the current density. 
Assuming the medium is dispersionless and source-free $J=0$, one can integrate Eqs. (\ref{eq1}) and find that the LHS of the equations vanish, which yields the conventional interface condition being discussed in the broad literature \cite{mendoncca2002time, xiao2014reflection, ruiz1978characteristics, auyeung1983phase, cirone1997photon}
\begin{equation}
    \begin{aligned}
\bm{B}(t_0^+)=\bm{B}(t_0^-),\\
\bm{D}(t_0^+)=\bm{D}(t_0^-).
    \end{aligned}\label{eq2}
\end{equation}
These continuity relations are referred to as temporal interface conditions in the following, which ensure the continuity of the magnetic flux density $\bm{B}$ and electric displacement $\bm{D}$ at a temporal interface, and further imply the continuity of electromagnetic (Minkowski) momentum density \cite{ortega2023tutorial},
\begin{equation}
    \bm{p}_{\mathrm{M}}=\bm{D}\times\bm{B},
\label{eq3}
\end{equation}
The electromagnetic momentum density represents the canonical momentum density of the electromagnetic wave, associated with the wavevector and spatial translation symmetry. 
In contrast, the Abraham momentum density $\bm{p}_{\mathrm{A}}=\mu_0\varepsilon_0\bm{E}\times\bm{H}.$ varies at a temporal interface, which is the kinetic momentum of the electromagnetic field, proportional to the Poynting vector, hence associated with energy flow. 
Therefore, an ideal temporal interface conserves momentum, while the energy flux and  electromagnetic energy can change.

We should emphasize that the conservation of electromagnetic momentum density can be broken due to the microscopic implementation of temporal interfaces \cite{galiffi2025electrodynamics}. 
In particular, the continuity of $\bm D$ assumes the absence of an impulsive source current. 
When additional charge is created or annihilated at a temporal interface, the interface condition is rewritten by
\begin{equation}
    \begin{aligned}
    \bm{B}(t_0^+)-\bm{B}(t_0^-)&=0,\\
    \bm{D}(t_0^+)-\bm{D}(t_0^-)&=\bm{\sigma}_{\mathrm{se}},
    \end{aligned}\label{eq4}
\end{equation}
where $\bm\sigma_{\mathrm{se}}=-\int^{t_0^+}_{t_0^-} \bm{J}_s(t) dt$, and $\bm{J}_s$ is the electric surface density at the temporal interface. 
If $\bm{\sigma}_{\mathrm{se}} \neq 0$, the electric displacement $\bm{D}$ is no longer conserved. 
This illustrates how different microscopic implementations of a temporal interface can lead to different interface conditions, even when they produce similar macroscopic changes.

We then consider a concrete example, where the permittivity of an isotropic nondispersive medium is changed from $\varepsilon_1$ to $\varepsilon_2$ at $t_0$. 
For simplicity, we consider a one-dimensional system with a fixed polarization, where $D$ denotes the corresponding scalar component of the electric displacement, and $x$ and $k$ represent the spatial coordinate and wavevector component, respectively.
The electric displacement field before the temporal interface is assumed to be a plane wave:
\begin{equation}
    D(x,t<t_0)=D_{\mathrm{i}}\exp{i[kx-\omega_1 (t-t_0)]}.
    \label{eq5}
\end{equation}
After the temporal interface, the total electric field is the superposition of the time-refracted wave and time-reflected wave:
\begin{equation}
    D(x,t>t_0)=D_{\mathrm{t}}\exp{i[kx-\omega_2 (t-t_0)]}+D_{\mathrm{r}}\exp{i[kx+\omega_2 (t-t_0)]},
    \label{eq6}
\end{equation}
where $D_{\mathrm{t}}$ and $D_{\mathrm{r}}$ denote the amplitude of the refracted wave and reflected wave, respectively. 
Note that the wavevector $k$ and frequency $\omega_i$ satisfy the dispersion relation $k=n_i \omega_i /c$, and hence we have the temporal Snell's law \cite{mendoncca2002time}
\begin{equation}
    n_1 \omega_1=n_2 \omega_2,
    \label{eq7}
\end{equation}
which gives the frequency conversion at a temporal interface. 
Applying the interface condition Eq. (\ref{eq2}), we have the following relations
\begin{equation}
    \begin{aligned}
    D_{\mathrm{i}}&=D_{\mathrm{t}}+D_{\mathrm{r}},\\
    \sqrt{\frac{\mu_0}{\varepsilon_1}}D_{\mathrm{i}}&=\sqrt{\frac{\mu_0}{\varepsilon_2}}(D_{\mathrm{t}}-D_{\mathrm{r}}).
    \end{aligned}\label{eq8}
\end{equation}
By substituting the constitutive relation between $D$ and $E$, we can obtain the temporal scattering coefficients
\begin{equation}
    \begin{aligned}
    T=\frac{E_t}{E_i}=\frac{1}{2} \left(\frac{\varepsilon_1}{\varepsilon_2}+\frac{\sqrt{\mu_1 \varepsilon_1}}{\sqrt{\mu_2 \varepsilon_2}} \right),\\
    R=\frac{E_r}{E_i}=\frac{1}{2}\left(\frac{\varepsilon_1}{\varepsilon_2}-\frac{\sqrt{\mu_1 \varepsilon_1}}{\sqrt{\mu_2 \varepsilon_2}}\right).
    \end{aligned}\label{eq9}
\end{equation}
We should emphasize that these coefficients are derived for an abrupt temporal interface under the assumption of spatial homogeneity, isotropy, and instantaneous material response, and may be modified under different conditions. 
Moreover, the temporal scattering is also expected to be changed when considering a more realistic or complicated material, such as anisotropy, dispersion, and broken spatial translational symmetry \cite{solis2021time, yin2022temporal, stefanini2023temporal, yang2023generalized, mostafa2024temporal, feng2024temporal, mirmoosa2024time, ptitcyn2025temporal}.

The above discussion concerns electromagnetic temporal interfaces governed directly by Maxwell's equations. A related but distinct framework is provided by systems described by effective Schr\"odinger-type equations, which are widely used in photonic lattice models and other engineered wave systems. For instance, under the paraxial approximation, the propagation coordinate plays the role of an effective evolution time, while the transverse refractive-index profile acts as a potential \cite{pertsch2002anomalous,christodoulides2003discretizing,szameit2010discrete,rechtsman2013photonic,xia2024deep}. Similar first-order evolution equations arise in coupled-mode descriptions and synthetic-dimension systems \cite{sounas2013giant,koutserimpas2018nonreciprocal,yuan2018synthetic,ozawa2019topological}. Such mappings depend on the underlying approximations and should not be regarded as a general property of optical systems. 
It is therefore useful to distinguish the interface conditions of these effective photonic models from those of a genuine electromagnetic temporal interface.
For a system governed by the Schr\"odinger equation
\begin{equation}
i\partial_t \left| \psi (t) \right>=H(t)\left| \psi (t) \right>,\label{eq10}
\end{equation}
when the Hamiltonian experiences a sudden jump at $t=t_0$, integrating Eq. (\ref{eq10}) over $t$ yields
\begin{equation}
\left|\psi(t_0^+) \right>=\left| \psi(t_0^-) \right>.\label{eq11}
\end{equation}
Such a interface condition indicates the continuity of the wave function at a temporal interface, while its first-order time derivatives can be discontinuous. 
The wave function is then expanded in the eigenbasis of the new Hamiltonian. 
Note that this projection generally does not generate a pair of time-refracted and time-reflected waves, as a system governed by Schr\"odinger equation (i.e., Schr\"odinger-type equation) does not intrinsically possess paired positive- and negative-frequency branches, like a Maxwell's-equation-governed systems (i.e., Maxwell-type equation). 
Nevertheless, by properly engineering Schr\"odinger-type systems, one can reproduce some features of electromagnetic temporal scattering \cite{long2023time, ye2023reconfigurable, dong2024quantum, jin2025temporal}. 
For example, a multiband Hamiltonian may be designed such that a sudden temporal change projects the initial state onto two bands with opposite group velocities, which can be identified as time-refracted and time-reflected wave packets \cite{dong2024quantum}. 
Such behavior originates from the engineered band structure rather than from an intrinsic positive-negative-frequency pairing.

The distinction between Schr\"odinger-type and electromagnetic temporal interfaces becomes particularly important when we deal with quantized electromagnetic fields. 
For a finite sudden change of the medium, the quantum state remains continuous across the temporal interface, i.e., $\left|\psi(t_0^+) \right>=\left|\psi(t_0^-) \right>$, as required by the Schr\"odinger equation. 
In contrast, the field operators must satisfy the interface conditions inherited from Maxwell's equations, namely the continuity of $\bm{D}$ and $\bm{B}$ for an ideal source-free case. 
Since the normal modes change across the interface, expressing these continuous field operators in the respective mode bases leads to a Bogoliubov transformation that mixes the annihilation operator of the $\bm{k}$ mode with the creation operator of the $-\bm{k}$ mode \cite{mendoncca2000quantum, kim2026quantum}. 
This transformation has the form of two-mode squeezing and preserves the bosonic commutation relations. 
A nonadiabatic temporal change can therefore convert vacuum fluctuations into correlated photon pairs with opposite momenta, reflecting momentum conservation in a quantum level \cite{mendoncca2005time}.

\subsection{Temporal scattering matrix and temporal slabs}
\begin{figure}[bp]
    \centering
    \includegraphics[width=8cm]{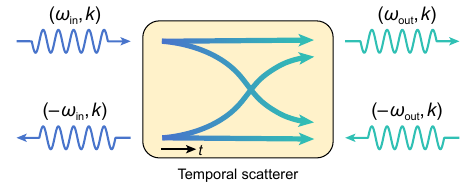}
    \caption{Temporal scattering process for a time-varying system. 
    A temporal scatterer maps the positive- and negative-frequency input modes onto the corresponding output modes. 
    The scatterer may represent a single temporal interface, a temporal slab, or a time-dependent structure.}\label{figure.2}
\end{figure}

Despite their different governing equations and temporal interface conditions, time-varying systems can generally be described within a modal scattering framework. We now consider a general temporal scatterer, as illustrated in Fig. \ref{figure.2}. 
Such a temporal scatterer can be a single temporal interface, a temporal slab, or a time-dependent structure, which convert the input field to the output field \cite{yin2023scattering},
\begin{equation}
\bm{c}_{\mathrm{out}}=S \bm{c}_{\mathrm{in}}.
\label{eq12}
\end{equation}
Here, $\bm{c}_{\mathrm{in}}$ ($\bm{c}_{\mathrm{out}}$) denotes the vector of input (output) modal amplitudes, whose elements represent the amplitude of each mode. 
$S$ is the temporal scattering matrix, and its element $S_{mn}$ gives the conversion from the $n$-th input mode to the $m$-th output mode. 
Note that this temporal scattering matrix is different from the conventional scattering matrix of a spatial structure. 
A spatial scattering matrix relates waves incident from different spatial ports to the corresponding outgoing waves. 
By contrast, a temporal scattering matrix maps all modal amplitudes before the temporal scattering onto the ones after, reflecting the causality of the evolution in time \cite{yin2023scattering}.

The explicit form of $S$ depends on the physical realization of the temporal scatterer. 
For a single temporal interface, it is determined by the change of modal basis together with the interface condition. 
We assume a temporal interface occurs at $t=t_0$, and define the modal matrices
\begin{equation}
    \begin{aligned}
V^{-}&=\left(\left|u_1^-\right>, \left|u_2^-\right>, \dots, \left|u_N^- \right>\right),\\
V^{+}&=\left(\left|u_1^+\right>, \left|u_2^+\right>, \dots, \left|u_N^+ \right>\right),
    \end{aligned}
    \label{eq13}
\end{equation}
where $\left|u_i^-\right>$ ($\left|u_i^+\right>$) is the eigenstate before (after) the temporal interface. 
The input and output states can then be written as
\begin{equation}
\begin{aligned}
\left|\psi_-\right>&=V^{-} \bm{c}^-,\\
\left|\psi_+\right>&=V^{+} \bm{c}^+.
\end{aligned}
\label{eq14}
\end{equation}
The interface condition at the temporal interface is expressed by
\begin{equation}
\left|\psi_+\right>=M_{\mathrm{B}}\left|\psi_-\right>,
\label{eq15}
\end{equation}
where $M_{\mathrm{B}}$ is determined by the specific interface condition. 
By substituting Eq. (\ref{eq14}) into Eq. (\ref{eq15}), one can obtain the temporal scattering matrix
\begin{equation}
S=\widetilde{V}^+ M_{\mathrm{B}} V^{-},
\label{eq16}
\end{equation}
where
\begin{equation}
\widetilde{V}^{\pm}=\left(
\left<u_1^{\pm}\right|,
\left<u_2^{\pm}\right|,
\cdots,
\left<u_N^{\pm}\right|\right)^T,
\label{eq17}
\end{equation}
being the dual modal matrix of $V^{\pm}$.
For continuous interface condition ($M_{\mathrm{B}}=I$), which is widely used in Schr\"odinger-type systems \cite{long2023time, ye2023reconfigurable, dong2024quantum}, the scattering matrix is given by
\begin{equation}
S=\widetilde{V}^+ V^-.
\label{eq18}
\end{equation}
The matrix elements of S are given by
\begin{equation}
S_{mn}=\left<u_m^+|u_n^-\right>,
\label{eq19}
\end{equation}
which yields the overlap between the eigenstates before and after the temporal interface. 
Moreover, as the eigenstates form orthonormal bases, the temporal scattering matrix is unitary $S^{\dagger}S=I$, hence preserving the total probability $\sum_n |c_n^+|^2=\sum_n |c_n^-|^2$. 
For a non-Hermitian Schr\"odinger-type system, the temporal scattering matrix is constructed by the overlap between the biorthogonal eigenstates \cite{brody2014biorthogonal}, $S_{mn}=\left<u_m^{\mathrm{L,+}}|u_n^{\mathrm{R,-}}\right>$, where the subscripts L and R denote the left and right eigenstates, respectively. 
Since the left and right eigenstates themselves are generally nonorthogonal, the projection maps a set of nonorthogonal eigenstates to another one. 
Therefore, $S$ is generally nonunitary under the ordinary Euclidean inner product and $|c_n^{\pm}|^2$ cannot be interpreted as conventional probabilities \cite{ashida2020non-hermitian}. 

For a one-dimensional Maxwell-type system, the electric displacement and magnetic induction can be combined into a classical state vector,
\begin{equation}
|\psi_{\pm}\rangle=
\begin{pmatrix}
D_{\pm}\\
B_{\pm}
\end{pmatrix}.
\end{equation}
The electromagnetic modes are generally not orthogonal under the ordinary Euclidean inner product, and the corresponding temporal scattering matrix is therefore not necessarily unitary in this representation. Importantly, such nonunitarity does not imply the absence of a conserved quantity. For a lossless and nondispersive electromagnetic medium undergoing an abrupt temporal change, the continuity of $D$ and $B$ at the temporal interface leads to the conversion between positive- and negative-frequency modes. Although electromagnetic energy is generally not conserved because of its exchange with the time-varying medium, the corresponding scattering matrix obeys a generalized conservation relation. In an appropriately normalized modal basis, it satisfies the pseudo-unitary condition \cite{ME_SE}
\begin{equation}
S^\dagger J S=J,
\end{equation}
where $J=\mathrm{diag}(1,-1)$, and the positive and negative signs correspond to the two frequency branches. Consequently, the normalized time-transmission $\mathcal T$ and time-reflection coefficients $\mathcal R$ satisfy $\mathcal T-\mathcal R=1$, rather than the conventional relation $\mathcal T+\mathcal R=1$ for unitary scattering. This illustrates how temporal modulation can produce amplification without material gain while conserving wave action, i.e., a photon-number-like quantity where positive- and negative-frequency modes contribute with opposite signs \cite{globosits2024pseudounitary}. In the quantum description, the corresponding transformation is closely related to Bogoliubov mixing between positive- and negative-frequency modes. 
This distinction in the conserved quantities also leads to distinct Floquet spectra of time-periodic systems. In particular, the norm conservation in Hermitian Schr\"odinger-type systems constrains the Floquet quasifrequencies to remain real, allowing frequency gaps, while the wave-action conservation in lossless Maxwell systems permits complex quasifrequencies and hence the formation of momentum gaps in PTCs.
More generally, the conservation relations satisfied by temporal scattering matrices can be understood from the mathematical structure of the underlying wave equations and the corresponding temporal interface conditions \cite{ME_SE}. Related pseudo-unitary relations have also been established for Floquet scattering matrices of spatially finite, time-periodic photonic media, providing a generalized framework for describing scattering between different Floquet channels \cite{globosits2024pseudounitary}. This framework has further been applied to lasing, and coherent perfect absorption in Floquet scattering systems \cite{globosits2026exceptional}.

\begin{figure}[!b]
    \centering
    \includegraphics[width=8cm]{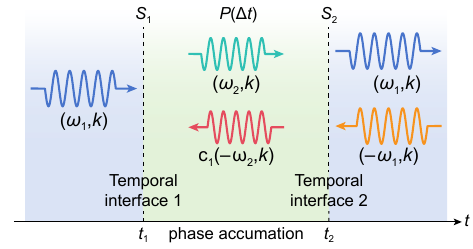}
    \caption{Illustration of a temporal slab and temporal interference. 
    An incident wave with frequency $\omega_1$ generates temporal interfered waves with frequencies $\omega_1$ and $-\omega_1$.}\label{figure.3}
\end{figure}

We then consider the case of a temporal slab composed of two temporal interfaces at $t_1$ and $t_2$ in Fig. \ref{figure.3}. 
The incident wave first projects into the new eigenbasis at $t_1$, and then each eigenstate evolves independently, as described by the evolution matrix
\begin{equation}
P(\Delta t)=\begin{pmatrix}
e^{-i\lambda_1 \Delta t} & 0 & 0 & 0\\
0 & e^{-i\lambda_2 \Delta t} & 0 & 0\\
0 & 0 & \ddots & 0\\
0 & 0 & 0 & e^{-i\lambda_N \Delta t}
\end{pmatrix},
\label{eq21}
\end{equation}
where $\Delta t=t_2-t_1$, and $\lambda_i$ is the eigenfrequency of the $i$-th eigenstate. 
Hence, these eigenstates accumulate different dynamical phases during evolution over the time interval $\Delta t$. 
Finally, the system returns to its original form, while the state is different from the incident one due to the accumulated phase difference in the temporal slab. 
The overall transfer matrix of the temporal slab is then written by
\begin{equation}
M_{\mathrm{slab}}=S_2 P(\Delta t) S_1,
\label{eq22}
\end{equation}
where $S_1$ and $S_2$ are the temporal scattering matrices at $t_1$ and $t_2$, respectively. 
Such a process can be regarded as an interference in the time domain. 
For a Maxwell-type system in Fig. \ref{figure.3}, the transmission and reflection amplitudes of the temporal slab are given by \cite{mendonca2003temporal}
\begin{equation}
    \begin{aligned}
        T_{\mathrm{slab}}&=\left[\cos(\omega_2 \tau)+\frac{i}{2\alpha}(1+\alpha^2)\sin(\omega_2 \tau)\right]e^{-i\omega_1\tau},\\
        R_{\mathrm{slab}}&=-\frac{i}{2\alpha}(1-\alpha^2)\sin (\omega_2 \tau)e^{i \omega_1 \tau},
    \end{aligned}
\end{equation}
where $\alpha=n_1/n_2$ and $\omega_i= kc/n_i$. 
One sees that the transmission and reflection amplitudes can be tuned by the duration of the temporal slab. 
The maximum of reflection amplitude $R$ occurs at $\omega_2 \tau$ equal to an odd multiple of $\pi/2$, which also gives the maximum of the output energy. 
Conversely, the reflection takes its minimum at $\omega_2 \tau$ equal to an even multiple of $\pi/2$, where the temporal slab becomes reflectionless. 
We can further cascade multiple temporal slabs into a temporal multilayered structure, and the transfer matrix is written by
\begin{equation}
M_{\mathrm{ML}}=S_{N+1}P_N(\Delta t_N)S_N\dots P_2(\Delta t_2)S_{2}P_1(\Delta t_1)S_1.
\label{eq24}
\end{equation}
Such a temporal structure provides more degrees of freedom than a single temporal slab, which allows a more flexible tailoring of the scattering response. 
As the number of temporal layers increases, the number of interfering pathways grows, giving access to more complex temporal filtering, frequency conversion, and amplification \cite{ramaccia2021temporal}.

\begin{figure}[!t]
    \centering
    \includegraphics[width=17cm]{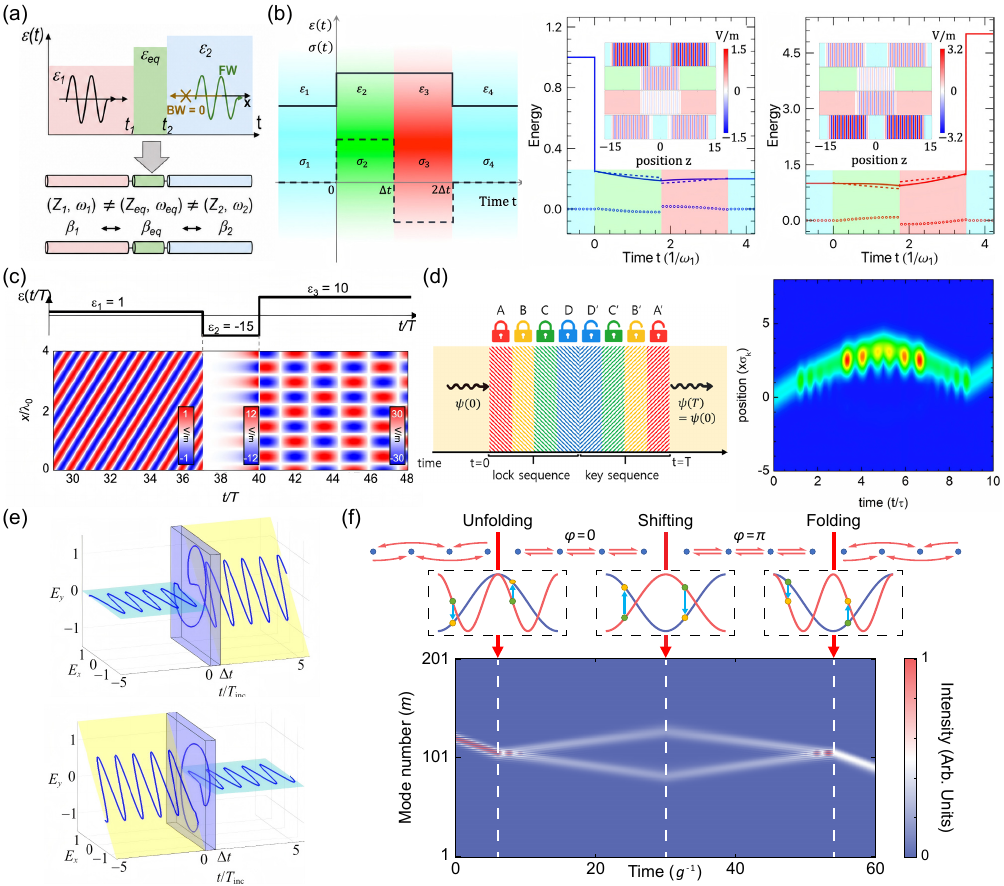}
    \caption{(a) Illustration of analog antireflection temporal coating \captioncite{pacheco2020antireflection}. 
    (b) Schematic of a temporal structure with temporal parity-time symmetry, and the time evolution of the normalized energy (solid lines) and its two components (dashed lines and empty circles) with different relative phases between the input waves \captioncite{li2021temporal}. (c) Wave freezing and exponential growth in a temporal non-foster metastructures \captioncite{pacheco2025holding}. 
    (d) Illustration of complete waveform restoration via temporal rewinding \captioncite{kim2025deterministic}. 
    (e) Evolution of a monochromatic plane wave and a $45^{\circ}$-polarized light in a chiral temporal slab, respectively \captioncite{yin2022temporal}. 
    (f) Simulating non-Abelian gauge field in a temporal multilayered structure \captioncite{dong2024temporal}.
    Panel (a) is reproduced from Ref.~\captioncite{pacheco2020antireflection}, \textcopyright~2020 Optical Society of America under the terms of theOSA Open Access Publishing Agreement; panel (b) is reproduced with permission from Ref.~\captioncite{li2021temporal}, \textcopyright~2021 American Physical Society; panel (c) is reproduced from Ref.~\captioncite{pacheco2025holding}, \textcopyright~The Author(s) 2025 (CC BY-NC-ND 4.0); panel (d) is reproduced from Ref.~\captioncite{kim2025deterministic}, \textcopyright~2025 The Authors (CC BY 4.0); panel (e) is reproduced from Ref.~\captioncite{yin2022temporal}, \textcopyright~2022 Optica Publishing Group under the terms of the Optica Open Access Publishing Agreement; and panel (f) is adapted with permission from Ref.~\captioncite{dong2024temporal}, \textcopyright~The Author(s) 2024.}\label{figure.4}
\end{figure}

Following this strategy, one can flexibly engineer the properties of waves through temporal scattering, phase accumulation, and temporal interference. 
The most direct example is the antireflection temporal coating in Fig. \ref{figure.4}(a), where a finite temporal layer is designed so that the backward-propagating waves generated at two temporal interfaces destructively interfere, suppressing temporal reflection while enabling frequency conversion of the forward-propagating wave \cite{pacheco2020antireflection}. 
The quantum counterpart of such coatings is further studied, which shifts quantum states in frequency while preserving photon statistics and avoiding thermal-noise amplification \cite{liberal2023quantum}. 
To realize a large frequency conversion, large refractive-index changes and substantial modulation energy are required, which is difficult for a single temporal slab in practice. 
Nevertheless, one can cascade multiple temporal operations to accumulate a large spectral shift \cite{apffel2022frequency}, which also relaxes the energy requirement for each temporal operation \cite{mai2022temporal}. 
Note that temporal modulation also exchange energy between the wave and the medium, allowing energy transformation. 
In particular, a temporal structure with temporal parity-time symmetry in Fig. \ref{figure.4}(b) manifests the dual phenomenon of laser-absorber pairs in its spatial counterpart \cite{chong2011pt}, where the output wave can be amplified or attenuated through tuning the relative phase between the input counterpropagating waves \cite{li2021temporal}. 
Another example is the temporal non-Foster metastructure shown in Fig. \ref{figure.4}(c) \cite{pacheco2025holding}. 
By switching the permittivity from a positive value to a dispersionless negative value and then switching it back, one can hold the wave in time, with its phase effectively frozen while its amplitude grows exponentially. 
Moreover, deterministic time rewinding can be realized from a carefully tailored modulation sequence [see Fig. \ref{figure.4}(d)], which reconstructs the amplitude and phase of the  initial wave \cite{kim2025deterministic}. 
Apart from frequency and energy control, temporal structure can manipulate the internal degree of freedom of light. 
For a chiral time-slab in Fig. \ref{figure.4}(e), an incident wave can split into circularly polarized components with different frequencies at the first temporal interface, then accumulate different phases within the temporal slab and finally interfere, producing a temporal analog of optical activity \cite{yin2022temporal}. 
In addition, an anisotropic temporal slab enables spin-dependent photonic operations, including spin-dependent analog computing and spin-orbit interaction effects for vortex generation \cite{rizza2023spin}. 
The same sequence of temporal switching, mode conversion, and phase accumulation can also be explored in effective Schr\"odinger-type systems, although the corresponding state variables and interface conditions differ from those of Maxwell's equations. 
For a pseudospin-1/2 Dirac equation, a temporal operation can be generated by switching the mass, Fermi velocity, and vector potential, giving rise to phenomena such as temporal Brewster effects, total interband transitions, and Dirac temporal crystals \cite{kim2023propagation}. 
In discrete physical systems, a sequence of temporal operations in Fig. \ref{figure.4}(f) can synthesize noncommuting operations on wave states, which emulates the effect of a non-Abelian gauge field, and further allows non-Abelian Aharonov-Bohm interference \cite{dong2024temporal}. 
From these examples, we can see that temporal slabs and multilayer structures are not merely time-domain analogs of spatial structures, but also offer a versatile platform for controlling multiple properties of light.

\subsection{Topological perspective of temporal interfaces}

\begin{figure}[!t]
    \centering
    \includegraphics[width=14.5cm]{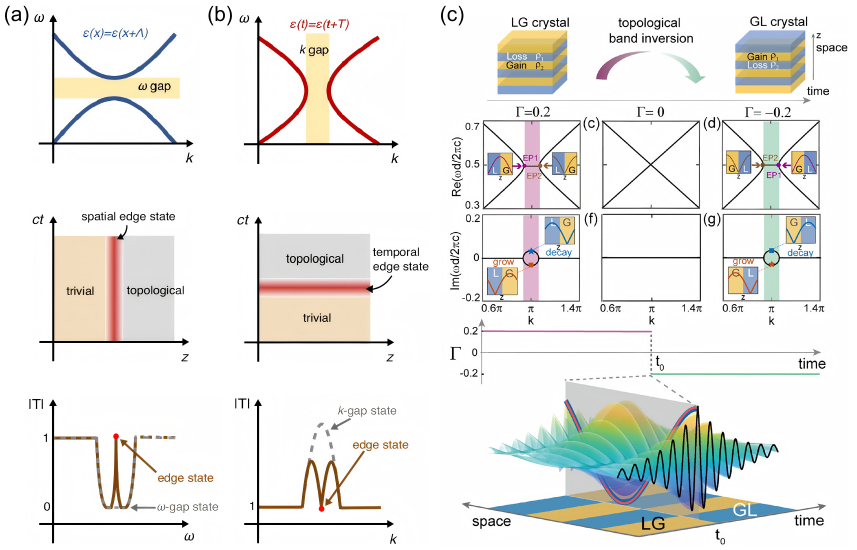}
    \caption{(a) Energy gap in a spatial photonic crystal, topological edge state, and transmission spectrum of a spatial topological domain wall \captioncite{xiong2025observation}. 
    (b) Momentum gap in a PTC, temporal topological edge state, and transmission spectrum of a temporal topological domain wall \captioncite{xiong2025observation}. 
    (c) Topological band inversion induced by non-Hermiticity $\Gamma$ flipping in a non-Hermitian spatial crystal, and the spatiotemporal profile of the temporal topological edge state localized at the temporal interface \captioncite{li2025topological}.
    Panels (a) and (b) are reproduced fromRef.~\captioncite{xiong2025observation},
\textcopyright~The Author(s) 2025 (CC BY-NC-ND 4.0);
panel (c) is reproduced with permission from
Ref.~\captioncite{li2025topological},
\textcopyright~2025 American Physical Society.}\label{figure.5}
\end{figure}

A temporal interface can be regarded as a sudden change of the band structure of the system. 
When the systems before and after the temporal interface carry different topologies, a temporal topological domain wall is created. 
A representative example is the temporal topological edge state in PTCs \cite{lustig2018topological}, reflecting a topological phase transition at the momentum gap, which is a temporal analog of a conventional spatial topological edge state [see Fig. \ref{figure.5}(a)]. 
To gain a better insight into the underlying physics, we first consider an effective non-Hermitian Dirac model
\begin{equation}
i\partial_t \left|\psi(t)\right> = H_{\mathrm{eff}}(q,t) \left|\psi(t)\right>, 
\label{eq25}
\end{equation}
with the effective Dirac Hamiltonian
\begin{equation}
H_{\mathrm{eff}}(q,t)=v q \sigma_z + i\mu(t) \sigma_x.
\label{eq26}
\end{equation}
Here, $q=k-k_0$ denotes the momentum deviation from the band-crossing point $k_0$, $v$ characterizes the slope of the linearly dispersing bands near the crossing, and $\sigma_x$ and $\sigma_z$ are Pauli matrices. The effective mass is purely imaginary, $m(t)=i\mu(t)$, where $\mu(t)$ is real.
The eigenfrequencies of the Hamiltonian are given by
\begin{equation}
\Omega_{\pm}(q)=\pm \sqrt{v^2 q^2 -\mu^2},
\label{eq27}
\end{equation}
and hence $\Omega_{\pm}(q)$ becomes purely imaginary for $|vq|<|\mu|$, leading to the opening of momentum gaps. 
A temporal interface between two momentum-gap topological phases can then be modeled as a temporal domain wall, where the imaginary mass flips sign
\begin{equation}
\mu(t<t_0)=-\mu(t>t_0).
\label{eq28}
\end{equation}
For a step-like temporal mass flip
\begin{equation}
\mu(t)=\mu_0 \hspace{0.1em} \mathrm{sgn}(t-t_0),
\label{eq29}
\end{equation}
one can solve the Dirac equation Eq. (\ref{eq25}) at the center of the momentum gap ($q=0$).
The solutions are given by
\begin{equation}
\left|\psi(t)\right>=e^{\pm \mu_0 |t-t_0|} \left(1,\pm  1\right)^T, 
\end{equation}
yielding a temporal localized (anti-localized) state, which exponentially grows (decays) before $t_0$, and then decays (grows) after $t_0$. 
Such a scenario is a time-domain analogue of the Jackiw-Rebbi bound state \cite{li2025topological}, where an imaginary mass flip occurs at a temporal domain wall instead of a real mass flip occurs at a spatial domain wall. 

This effective model captures the emergence of temporally localized topological states through a sign reversal of the imaginary Dirac mass across a temporal interface [see Fig. \ref{figure.5}(b)]. 
For example, in a non-Hermitian spatial crystal, a spatially periodic gain-loss profile opens a momentum gap, and reversing this profile changes the sign of the imaginary Dirac mass, giving rise to a temporal topological boundary state in Fig. \ref{figure.5}(c) \cite{li2025topological}. 
The same physical picture also applies to PTCs, whose momentum-band topology can be characterized by a Zak phase \cite{lustig2018topological}. 
A temporal interface between two PTCs with distinct topological phases can therefore support a temporal topological boundary state. 
Such states have been experimentally demonstrated in synthetic lattices \cite{ren2025observation}, acoustic cavity-tube structures \cite{tong2025observation}, and microwave transmission lines \cite{xiong2025observation}.
As a sidenote, one may notice that a temporal topological domain wall can support a localized state and an anti-localized one, while only the localized states are demonstrated in experiments. 
The reason is that a single incident wave in a spatially uniform time-periodic medium must obey momentum conservation, which naturally selects the gain mode. 
Nevertheless, a recent experiment shows that the decay mode can be excited by a carefully phase- and amplitude-matched pair of counterpropagating input waves, suggesting a possible route to experimentally assess the anti-localized state \cite{galiffi2026optical}.

\begin{figure}[!b]
    \centering
    \includegraphics[width=12.5cm]{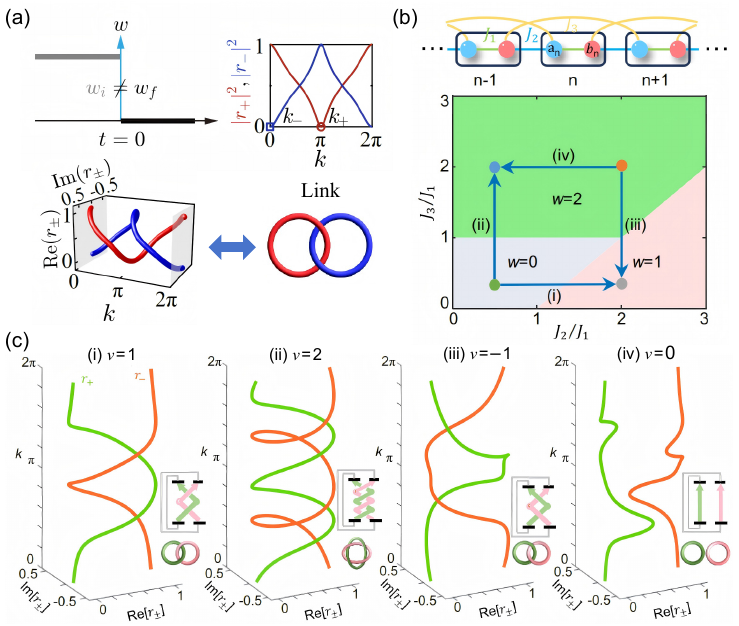}
    \caption{(a) When the topological invariant of the system changes at a temporal interface, the zeros of the time reflection and refraction coefficients corresponds to the bandgap-closing degeneracies at a temporal topological transition. 
    The evolution of time refraction and reflection coefficients versus $k$ forms a topological link \captioncite{xu2025probing}. 
    (b) Phase diagram of a long-range SSH model \captioncite{li2026observation}. 
    (c) For different temporal interfaces in (b), the time refraction and reflection coefficients form different topological links in the momentum space \captioncite{li2026observation}.
    Panel (a) is reproduced from Ref.~\captioncite{xu2025probing},
published by the American Physical Society under the terms
of the Creative Commons Attribution 4.0 International License
(CC BY 4.0);
panels (b) and (c) are reproduced from
Ref.~\captioncite{li2026observation},
\textcopyright~The Author(s) 2026 (CC BY-NC-ND 4.0).}\label{figure.6}
\end{figure}

The scattering at temporal interfaces can also be a probe of bulk band topology. 
Unlike the momentum-gap case where a temporal interface between two topological phases can support a temporally localized boundary state. Here, the topological transition of the system is encoded in the temporal scattering coefficients $S_{mn}=\left<u^+_m|u^-_n\right>$. 
For a two-band model, the time refraction and reflection coefficients $|r_+|$ and $|r_-|$ can vanish at certain $k$, which is the bandgap-closing degeneracies associated with the topological phase transition across the temporal interface [see Fig. \ref{figure.6}(a)] \cite{xu2025probing}. 
Moreover, the braiding of $|r_+|$ and $|r_-|$ in the momentum space forms a link with its linking number $\mathcal{L}$ satisfying
\begin{equation}
    \mathcal{L} =w_- -w_+,
    \label{eq31}
\end{equation}
where $w_-$ and $w_+$ are the winding numbers before and after the temporal interface, respectively. 
In this way, the scattering at temporal interface converts the difference between bulk topological invariants into measurable scattering coefficients, which is recently demonstrated in experiment [see Figs. \ref{figure.6}(b)-(c)] \cite{li2026observation}. 
In two-dimensional systems, the same idea extends beyond simple Chern-number classification. 
Even when two phases share the same Chern number, their Bloch wave-function geometry may remain distinct, and this hidden distinction appears as zeros of wave-function overlaps at a temporal topological phase transition \cite{wu2024edge}.
In non-Hermitian systems, the temporal scattering coefficients are replaced by a biorthogonal overlap between left and right eigenvectors \cite{jiang2026topological}. 
Their braiding still reflects the difference of winding numbers across the temporal interface. 
However, unlike the Hermitian case, due to the nonorthogonality of the post-interface eigenstates, the sum of the modal intensities is generally not conserved and cannot be directly interpreted as a conventional probability. 
Moreover, the same temporal interface can display propagation asymmetry, where reversing the direction of parameter switching changes attenuation of states into enhancement. 
A related but distinct case arises when the temporal interface crosses a parity-time symmetry phase transition, where the pre- and post-interface eigenvector geometry results in nearly momentum independent scattering coefficients, yielding broadband-stable temporal splitting \cite{jin2025momentum}. 

\section{Time-periodic system}

\subsection{Floquet theory}
The finite temporal scatterers discussed in the previous section provide the building blocks for more general time-periodic structures. 
When the material parameters or effective Hamiltonian are modulated periodically in time, the wave undergoes repeated scattering and phase accumulation over modulation cycles. 
Such systems are naturally described by Floquet theory, which plays the role of Bloch theory in the time domain, providing a framework for describing the long-time behavior of the system \cite{bukov2015universal, rodriguez2018floquet, oka2019floquet}. Before going into details, we clarify the terminology used throughout this section by distinguishing three closely related descriptions of temporal dynamics. The time-evolution operator $U(t_1,t_2)$ propagates a state from $t_1$ to $t_2$. For a time-periodic system, its one-period form $U_{\mathrm{F}}(t_0)\equiv U(t_0,t_0+T)$ is referred to as the Floquet operator, whose eigenvalues determine the Floquet quasifrequencies. The temporal transfer matrix describes the evolution in a chosen basis of field or modal amplitudes, typically constructed from successive temporal interfaces and propagation segments. Over a complete period, the transfer matrix and the Floquet operator represent the same evolution and are related by a change of basis. In the following, we first introduce the Floquet description and then discuss its transfer-matrix representation.

We consider a Schr\"odinger equation with a time-periodic Hamiltonian
\begin{equation}
    \begin{aligned}
    i\partial_t \left|\psi(t)\right>&=H(t) \left|\psi (t)\right>,\\
    H(t)&=H(t+T),
    \end{aligned}
\label{eq32}
\end{equation}
where $T=2\pi/\Omega$ is the time period and $\Omega$ is the modulation frequency. 
Here, the Schr\"odinger-type equation serves as a general mathematical representation of first-order wave evolution, rather than implying a specific physical implementation. Depending on the system, the Hamiltonian may be derived directly from Maxwell's equations or obtained as an effective description of coupled modes or synthetic lattices. 
Although the system breaks continuous time-translational symmetry, it still holds a discrete time-translational symmetry. 
Therefore, a solution of Eq. (\ref{eq32}) can be written in the Floquet form
\begin{equation}
\begin{aligned}
\left|\psi_{\alpha}(t)\right>&=e^{-i\varepsilon_{\alpha} t}\left|\Phi_{\alpha} (t)\right>,\\
 \left|\Phi_{\alpha} (t)\right>&=\left|\Phi_{\alpha}(t+T)\right>,
\end{aligned}
\label{eq33}
\end{equation}
where $\varepsilon_\alpha$ is the Floquet quasifrequency defined modulo
$\Omega$, and $\alpha$ labels the Floquet mode.
The origin of the quasifrequency can be seen more explicitly by rewriting the eigenequation in the extended Floquet-Hilbert space. 
Since both $H(t)$ and $\left|\Phi\right>$ are periodic, they can be expanded as
\begin{equation}
\begin{aligned}
H(t)&=\sum_n H_n e^{-in\Omega t},\\
\left|\Phi_{\alpha}(t)\right>&=\sum_n \left|\phi_{\alpha,n}\right> e^{-in\Omega t}.
\end{aligned}
\label{eq34}
\end{equation}
Substituting Eq. (\ref{eq34}) into Eq. (\ref{eq32}) maps a time-periodic problem onto a time-independent eigenvalue problem in an enlarged Hilbert space labelled by the Fourier mode index $m$
\begin{equation}
\sum_m \left(H_{n-m}-m\Omega\,\delta_{nm}I\right)
\left|\phi_{\alpha,m}\right\rangle
=\varepsilon_\alpha\left|\phi_{\alpha,n}\right\rangle,
\label{eq35}
\end{equation}
or equivalently
\begin{equation}
\begin{pmatrix}
\ddots & \vdots &\vdots &\vdots & \ddots\\[1mm]
\ldots & H_{0}+\Omega I & H_{-1} & H_{-2} & \cdots \\[1mm]
\ldots & H_{1} & H_{0} & H_{-1} & \cdots \\[1mm]
\ldots & H_{2} & H_{1} & H_{0}-\Omega I & \cdots \\[1mm]
\ddots & \vdots &\vdots &\vdots & \ddots
\end{pmatrix}
\begin{pmatrix}
    \vdots \\[1mm]
\left|\phi_{\alpha,-1}\right>\\[1mm]
\left|\phi_{\alpha,0}\right>\\[1mm]
\left|\phi_{\alpha,1}\right>\\[1mm]
\vdots
\end{pmatrix}=
\varepsilon_{\alpha} 
\begin{pmatrix}
\vdots \\[1mm]
\left|\phi_{\alpha,-1}\right>\\[1mm]
\left|\phi_{\alpha,0}\right>\\[1mm]
\left|\phi_{\alpha,1}\right>\\[1mm]
\vdots
\end{pmatrix}.
\label{eq36}
\end{equation}
If $\varepsilon_{\alpha}$ is the eigenvalue of Eq. (\ref{eq36}), $\varepsilon_{\alpha}+m\Omega$ can also satisfy Eq. (\ref{eq36}), yielding the periodicity of Floquet quasifrequency. 
This representation gives a simple physical picture: the periodically driven system is mapped to a synthetic lattice along the Floquet harmonic index $m$ \cite{oka2019floquet}. 
Each diagonal block $H_0-m\Omega I$ is a replica of the time-independent Hamiltonian $H_0$ shifted by $m\Omega$, while the Fourier component $H_{l}$ couples the replicas separated by $l$ harmonics. 

The Floquet-Hilbert space also provides a useful way to understand how an effective Floquet Hamiltonian emerges. 
Alternatively, the effective Hamiltonian and Floquet quasifrequencies can be obtained directly from the one-period evolution operator,
\begin{equation}
U_{\mathrm{F}}(t_0)=e^{-iH_{\mathrm{F}} T}.
\label{eq37}
\end{equation}
$H_{\mathrm{F}}$ can be interpreted as an effective Hamiltonian, which obtained by block-diagonalizing Eq. (\ref{eq36}) recast to treat interlayer transitions as virtual processes \cite{oka2019floquet}. 
For practical calculations, a convenient way to obtain the Floquet quasifrequency and the effective Hamiltonian is to work with the time-evolution operator over a period $T$
\begin{equation}
U_{\mathrm{F}}(t_0)=\mathcal{T} \exp \left[ -i \int^{t_0+T}_{t_0} H(t) dt \right],
\label{eq38}
\end{equation}
where $\mathcal{T}$ denotes time ordering. 
The effective Hamiltonian is then given by
\begin{equation}
H_{\mathrm{F}}=\frac{i}{T} \ln U_{\mathrm{F}}(t_0),
\label{eq39}
\end{equation}
and the Floquet eigenvalue problem is written as
\begin{equation}
U_{\mathrm{F}}(t_0) \left|\psi_{\alpha}(t_0)\right>=\lambda_{\alpha} \left|\psi_{\alpha}(t_0)\right>,
\label{eq40}
\end{equation}
with $\lambda_{\alpha}=e^{-i\varepsilon_{\alpha}T}$. 

\subsection{Temporal transfer-matrix method and momentum-eigenvalue reformulation}
The temporal transfer-matrix method provides an equivalent representation of wave evolution in a periodically modulated medium. As introduced in Eq. (\ref{eq24}), the transfer matrix describes the evolution of modal amplitudes across successive temporal interfaces and propagation segments. Using the modal matrix $V$, the one-period transfer matrix is related to the Floquet operator by $M(T)=V^{-1}U_{\mathrm{F}}(t_0)V$. Thus, the two operators are related by a similarity transformation and share the same eigenvalues, yielding identical Floquet quasifrequencies.
In the conventional formulation, the eigenvalue problem is written as
\begin{equation}
M(k, T)\left|\psi\right>=e^ {-i \phi(k)}\left|\psi\right>,
\label{eq41}
\end{equation}
where $M(k, T)$ is the temporal transfer matrix over one period, and $\phi(k)=\varepsilon(k) T$ is the Floquet phase accumulated over one modulation period. 
The propagation phase in the $j$-th time segment is $\beta_j(k)=\omega_j(k)\Delta t_j$, which is generally different from one segment to another. 
Therefore, the Floquet quasifrequency is obtained from the eigenvalues of $M(k,T)$ at a given wavevector $k$. 

A different formulation becomes possible when the propagation phase is identical in all time segments \cite{yang2025topologically, ni2025topological, he2026temporal}
\begin{equation}
\beta_j(k)=\omega_j(k)\Delta t_j=\beta,
\label{eq42}
\end{equation}
which can be realized by appropriately choosing the segment durations and material parameters. 
This condition allows the segment-dependent propagation phases to collapse into a single spectral variable $\beta$, enabling the reformulation of the eigenvalue problem. 
One can define a state vector $\left|u\right>$ composed of the field amplitudes $\left|u_i\right>$ at each temporal interface within one temporal unit cell
\begin{equation}
\left|u\right>=(\left|u_1\right>, \left|u_2\right>, \dots, \left|u_N\right>)^T.
\label{eq43}
\end{equation}
By applying the transfer matrix across each time segment and imposing the Floquet condition, the resulting eigenvalue equation generally takes the form
\begin{equation}
\mathcal{U}(\phi)\left|u\right>=e^{-i\beta}\left|u\right>.
\label{eq44}
\end{equation}
We emphasize that $\mathcal U(\phi)$ denotes an auxiliary operator in the reformulated eigenvalue problem, rather than the original Floquet operator. 
One sees that the Floquet phase $\phi$ becomes a Bloch-like or synthetic momentum in the reformulated eigenvalue problem, whereas the propagation phase $\beta$ becomes an eigenvalue. 
This reformulation converts the temporal transfer problem into an effective lattice or band eigenproblem, revealing the hidden topology of temporal structures \cite{yang2025topologically, ni2025topological, he2026temporal}. 
For a binary temporal modulation, the two segments of the temporal unit cell act as two sublattice degrees of freedom. 
The reformulated eigenequation takes an SSH-like form with $\phi$ playing the role of the one-dimensional Bloch momentum. 
The resulting chiral symmetry enables the definition of a quantized winding number and topologically protected temporal edge states \cite{yang2025topologically}. 
The same formulation can be extended by introducing additional modulation parameters as momenta reciprocal to a synthetic space, and hence topological structures in a higher-dimensional synthetic space can be defined, such as Chern number in photonic time quasicrystals \cite{ni2025topological} and a Weyl semimetal in anisotropic PTCs \cite{he2026temporal}.

\subsection{Photonic time crystals: fundamentals}

\begin{figure}[!t]
    \centering
    \includegraphics[width=8cm]{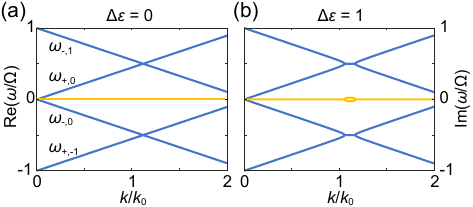}
    \caption{(a) Real part (blue lines) and imaginary part (yellow lines) of Floquet Bands for a PTC with $\mathrm{\Delta} \varepsilon \rightarrow 0$ and (b) $\mathrm{\Delta} \varepsilon \rightarrow 0$ respectively, where $\bar{\varepsilon}=5$, $k_0=2\pi/Tc$.}\label{figure.7}
\end{figure}

We now consider a representative example of time-periodic systems, i.e., PTCs. 
Such a temporal structure is spatially homogeneous while its constitutive parameters undergo periodic temporal modulation. 
Consider a simple case, where a spatially homogeneous, isotropic, and dispersionless medium has time-periodic permittivity $\varepsilon(t)=\bar{\varepsilon}[1+\Delta \varepsilon \cos (\Omega t)]$ \cite{wang2018photonic}. 
Without loss of generality, we consider a wave propagating along the $x$ direction with the electric field and magnetic field in the $y$ and $z$ directions, respectively. 
One can rewrite Maxwell's equation as
\begin{equation}
i\partial_t \begin{pmatrix}
D\\B
\end{pmatrix}=H_{\mathrm{PTC}}(k)\begin{pmatrix}
D\\B
\end{pmatrix}=k\begin{pmatrix}
0 & 1\\
\varepsilon^{-1} (t) & 0
\end{pmatrix}
\begin{pmatrix}
D\\B
\end{pmatrix}.
\label{eq45}
\end{equation}
Note that even when $\varepsilon$ is purely real, $H_{\mathrm{PTC}}(k)$ is non-Hermitian, which does not arise from material gain or loss. 
For a static medium, i.e., $\Delta \varepsilon = 0$, $H_{\mathrm{PTC}}(k)$ can be converted into a Hermitian form by a similarity transformation \cite{wang2018photonic}
\begin{equation}
\tilde{H}(k)=G^{-1}H_{\mathrm{PTC}}(k)G,
\label{eq46}
\end{equation}
with
\begin{equation}
G=\begin{pmatrix}
    i \sqrt{\bar{\varepsilon}} & -i \sqrt{\bar{\varepsilon}}\\
    1 & 1
\end{pmatrix}.
\label{eq47}
\end{equation}
One see that the Hamiltonian $\tilde{H}(k)$ has two eigenvalues $\omega_{\pm,0}(k)=\pm k/\sqrt{\bar{\varepsilon}}$, corresponding to the positive and negative frequency branches. 
The periodic modulation produces Floquet replicas $\omega_{\pm, n}(k)=\omega_{\pm,0}(k)+n\Omega$ as illustrated in Fig. \ref{figure.7}(a), where different orders of Floquet replicas form band-crossing points. 
The lowest-order crossing point relevant to the first momentum gap occurs at $\omega_{-, 1}(k_{\mathrm{cross}})=\omega_{+, 0}(k_{\mathrm{cross}})=\Omega/2$. 
Thus, the center of the first momentum gap is fixed by a parametric resonance between a positive-frequency mode and a shifted negative-frequency mode.

When $\varepsilon$ is modulated, one can still perform the same similarity transformation in Eq. (\ref{eq47}), and then obtain a time-dependent Hamiltonian $\tilde{H}(k,t)$, which can be rewritten in the form of the Floquet Hamiltonian in Eq. (\ref{eq36}). 
Under weak modulation $\Delta \varepsilon \ll 1$ condition, at the band crossing point, we can simply consider the interaction between the resonant bands $\omega_{-, 1}(k)$ and $\omega_{+, 0}(k)$, and neglect the effect of the higher-order components $H_n$ ($|n|>1)$, leading to a reduced Hamiltonian
\begin{equation}
\tilde{H}_{\mathrm{red}}=\begin{pmatrix}
-\frac{k-k_{\mathrm{cross}}}{\sqrt{\bar{\varepsilon}}} & g \\
-g & \frac{k-k_{\mathrm{cross}}}{\sqrt{\bar{\varepsilon}}}
\end{pmatrix}+\frac{\Omega}{2}I,
\label{eq48}
\end{equation}
where $g=\Delta \varepsilon \Omega/8$, denoting the coupling strength between Floquet replicas. 
Note that the coupling between these Floquet replicas is non-Hermitian, originating from the energy exchange between the modulation source and light field, which opens the momentum gap
\begin{equation}
\omega=\pm\sqrt{(k-k_{\mathrm{cross}})^2/\bar{\varepsilon}-g^2}+\frac{\Omega}{2}.
\label{eq49}
\end{equation}
When $|(k-k_{\mathrm{cross}})/\sqrt{\bar{\varepsilon}}| >|g|$, the values of Floquet quasifrequency $\omega$ are real numbers, and the corresponding modes hold an oscillating feature during evolution. 
However, for $|(k-k_{\mathrm{cross}})/\sqrt{\bar{\varepsilon}}| < |g|$, $\omega$ forms a complex-conjugate pair. 
One mode undergoes an exponential amplification, whereas the other one decays. 
The change of Floquet quasifrequency suggests a parity-time symmetry transition along the $k$ axis. 
At the critical points $|(k-k_{\mathrm{cross}})/\bar{\varepsilon}| = |g|$, both quasifrequencies and eigenvectors coalesce, which are referred to as exceptional points. 
The same transition appears in the transfer matrix of the temporal unit cell
\begin{equation}
M(k,T)\left|u_{\pm}(k, 0)\right>=\lambda_{\pm}(k)\left|u_{\pm}(k, 0)\right>=e^{-i\omega_{\pm}(k)T}\left|u_{\pm}(k, 0)\right>.
\label{eq50}
\end{equation}
In lossless cases, we have $\mathrm{det}M(k,T)=1$, so $\lambda_{+}(k)$ and $\lambda_{-}(k)$ form a reciprocal pair. 
The dispersion relation is given by
\begin{equation}
\cos(\omega_{\pm}T)=\frac{1}{2}\mathrm{Tr} M(k,T).
\label{eq51}
\end{equation}
For the lossless two-mode model considered here, the one-period transfer matrix is unimodular and has a real trace. Therefore, $|\mathrm{Tr}[M(k,T)]|<2$ corresponds to Floquet multipliers on the unit circle, whereas $|\mathrm{Tr}[M(k,T)]|>2$ gives a reciprocal pair with $|\lambda_+(k)|>1$ and $|\lambda_-(k)|<1$, corresponding to momentum-gap modes.

The transfer-matrix picture also establishes a direct connection between the interference effects of a single temporal slab and the formation of momentum gaps in PTCs. To illustrate this connection more explicitly, we consider a binary PTC, where the refractive index changes between $n_1$ and $n_2$ over durations $\tau_1$ and $\tau_2$, respectively. Defining the accumulated phases as $\phi_j=\omega_j(k)\tau_j$, the Floquet dispersion relation becomes \cite{hwang2024waveform}
\begin{equation}
\cos(\omega_F T)=\cos\phi_1\cos\phi_2-\frac{1}{2}\left(\frac{Z_1}{Z_2}+\frac{Z_2}{Z_1}\right)\sin\phi_1\sin\phi_2,
\end{equation}
where $T=\tau_1+\tau_2$ and $Z_j=\sqrt{\mu_j/\varepsilon_j}$ denotes the wave impedance of each medium. Similar to the constructive-interference condition of a single temporal slab, the magnitude of the Floquet growth factor is maximized when both accumulated phases are odd multiples of $\pi/2$, for a fixed impedance contrast. In particular, $\phi_1=\phi_2=\pi/2$ requires $\tau_1/\tau_2=n_1/n_2$, which corresponds to a duty cycle $\tau_1/T=n_1/(n_1+n_2)$. For a given wavevector, the total modulation period further determines whether the quarter-wave condition is satisfied. In contrast, when either accumulated phase is an integer multiple of $\pi$, no momentum gap opens at that wavevector. Thus, the interference in a single slab also provides an intuitive picture of how periodic temporal modulation produces Floquet instabilities \cite{asgari2024theory}.

In realistic PTCs, however, modulation-induced amplification competes with material absorption. The resulting net growth or decay is determined by the Floquet multipliers of the complete system, $\lambda_\alpha=e^{-i\omega_\alpha T}$. A Floquet mode exhibits net exponential amplification only when $|\lambda_\alpha|>1$, or equivalently $\mathrm{Im}(\omega_\alpha)>0$. Thus, the existence of a momentum gap in the corresponding lossless system does not necessarily guarantee amplification in the presence of material losses. Dissipation can also modify the complex Floquet spectrum and its associated exceptional points, although these spectral features may persist even when net amplification is suppressed \cite{park2022revealing,sun2025formulation}. This competition is particularly relevant at optical frequencies, where strong temporal modulation is often accompanied by significant material absorption \cite{guo2026plasmonic}.

The exponential amplification associated with momentum gaps in PTCs is closely related to conventional optical parametric amplification, as both involve parametric coupling and energy transfer from an external drive. However, the distinction between these phenomena has been discussed from different physical perspectives \cite{lyubarov2022amplified, sharabi2022spatiotemporal,khurgin2024photonic,asgari2024theory}. Conventional traveling-wave optical parametric amplification is typically described as a spatial amplification process, in which the coupling between signal and idler waves is governed by energy and momentum matching with the pump, with amplification accumulating over the interaction length. In contrast, a spatially homogeneous PTC conserves wavevector and supports a Floquet momentum gap, within which exponentially growing and decaying solutions exist. The growth is governed by the Floquet spectrum and the duration of temporal modulation. This distinction has motivated the interpretation of PTC amplification as nonresonant, in the sense that a finite range of wavevectors can be amplified without requiring exact phase matching for each mode \cite{lyubarov2022amplified,sharabi2022spatiotemporal}. Nevertheless, the momentum gap itself is associated with parametric coupling between Floquet harmonics and, in the weak-modulation limit, typically opens around a parametric resonance \cite{asgari2024theory}. Furthermore, the initial and boundary conditions play an essential role in determining how the growing and decaying solutions are excited in finite systems \cite{khurgin2024photonic}. Thus, we regard PTC amplification as a particular manifestation of parametric amplification, whose characteristic features arise from temporal periodicity and the associated scattering configuration.

The quantum interpretation of a photonic time crystal follows naturally from the temporal-interface picture discussed in Section 2.2. 
Each temporal interface mixes the annihilation operator of a mode with momentum $k$ with the creation operator of its counterpropagating partner with momentum $-k$. 
Periodic modulation further extends this elementary process to momentum-resolved Floquet dynamics \cite{kim2026quantum}. 
In the momentum bands, the resulting squeezing and photon population remain oscillatory, reflecting interference among pair-creation amplitudes generated at successive temporal interfaces. 
Inside a momentum gap, the squeezing parameter and vacuum-generated photon-pair population grow exponentially, which leads to an exponential enhancement of dynamical Casimir processes \cite{sustaeta2025quantum}. 
Moreover, the mean number of photon pairs generated through such processes coincides with the classical reflectivity of a PTC. 
Therefore, the classical amplification of a momentum gap can be interpreted as the amplification of vacuum fluctuations into correlated photon pairs with opposite momenta from the quantum perspective. 
In addition, a recent study suggests that the classical nonorthogonality of Floquet modes, quantified by the Petermann factor, provides a complementary measure of the underlying quantum squeezing \cite{kim2026classical}.

\begin{figure}[!t]
    \centering
    \includegraphics[width=8cm]{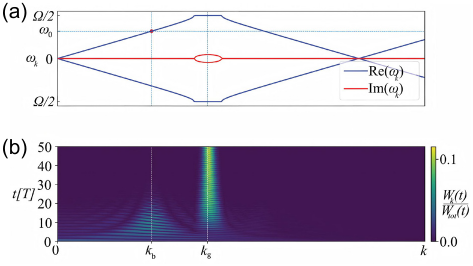}
    \caption{(a) Band structure of a PTC \captioncite{lyubarov2022amplified}. (b) Evolution of power spectrum of dipole emission versus wave number \captioncite{lyubarov2022amplified}.
    Panels (a) and (b) are reproduced with permission
from Ref.~\captioncite{lyubarov2022amplified},
\textcopyright~2022 The Authors,
exclusive licensee American Association for
the Advancement of Science.}\label{figure.8}
\end{figure}

The momentum gap not only governs the evolution of freely propagating waves, but also reshapes radiation generated inside a PTC. 
As shown in Fig. \ref{figure.8}, a dipole initially excites a broad range of momenta, while the component lying in the center of the momentum gap gradually becomes dominant, resulting in the narrowing of emission power spectrum \cite{lyubarov2022amplified}. 
This can be explained by the imaginary part of the Floquet quasifrequency, which determines the growth rate and reaches its maximum at the center of the momentum gap.
Therefore, the fastest-growing mode dominates the long-time radiation, and eventually produces monochromatic radiation spectrum. 
This mechanism provides a way to achieve spectrally selective emission, pointing towards a nonresonant tunable PTC laser. 
Furthermore, PTCs can also modify spontaneous emission from quantum emitters. 
Near the exceptional points at the momentum-gap edges, the Floquet modes become strongly nonorthogonal, so the emission rate cannot be inferred from the dispersion slope alone. 
Rather than vanishing at the gap edge, the spontaneous-emission decay rate can remain finite and be enhanced \cite{park2025spontaneous, lee2026analogs}. 
When modulation-induced amplification exceeds intrinsic loss, a nonequilibrium process can occur, where a ground-state emitter becomes excited while simultaneously emitting a photon. 
These effects illustrate that PTCs do not simply change the density of photonic states, but create a nonequilibrium electromagnetic environment, thus reshaping light-matter interactions.

\subsection{Photonic time crystals: beyond simple modulation}

With the basic picture of PTCs in mind, we now consider three kinds of generalizations, including modifying the modulation pattern, engineering material responses, and introducing additional physical mechanisms.

\begin{figure}[!b]
    \centering
    \includegraphics[width=17cm]{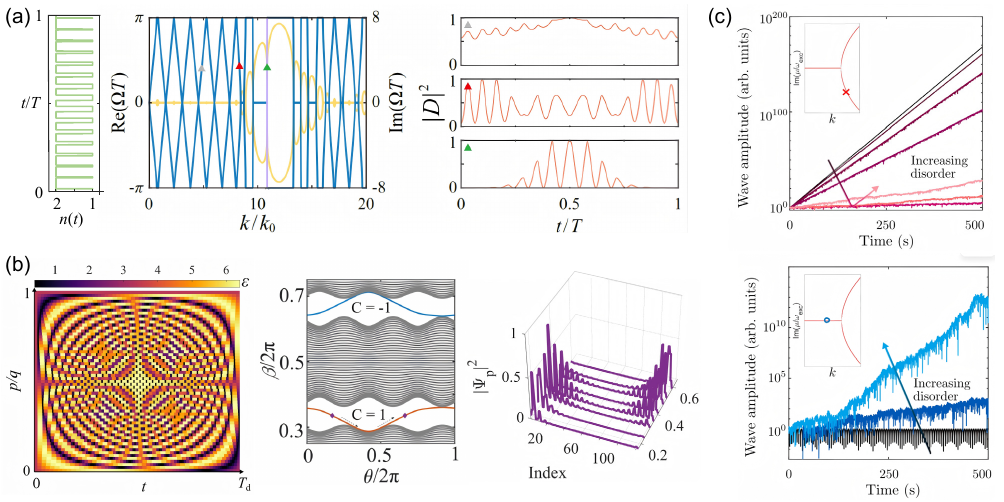}
    \caption{(a) Modulation pattern, band structure of a moir\'e PTC, and the intensity distribution of the modes at different momenta \captioncite{dong2025extremely}. 
    (b) Variation of the permittivity in the time domain for a photonic time quasicrystal with different configurations controlled by the parameters $p/q$, band structure along the synthetic momentum $\theta$, and topological pumping of temporal edge state by varying $\theta$ \captioncite{ni2025topological}. 
    (c) Growth of the wave amplitude versus time with different strengths of temporal disorder inside and outside a momentum gap, respectively \captioncite{apffel2022experimental}.
    Panels (a), (b), and (c) are reproduced
with permission from
Refs.~\captioncite{dong2025extremely},
\captioncite{ni2025topological}, and
\captioncite{apffel2022experimental},
respectively,
\textcopyright~2025, 2025, and 2022
American Physical Society.}\label{figure.9}
\end{figure}

The temporal modulation can be structured by introducing incommensurate periods, aperiodic order, or random fluctuations, providing direct control over Floquet interference during evolution. 
For example, superposing two binary temporal modulations with slightly different periods creates a moir\'e superlattice in time as illustrated in Fig. \ref{figure.9}(a) \cite{dong2025extremely}. 
The interference between two incommensurate temporal structures produces extremely narrow bands, which may enhance the sensitivity to perturbations near the band edge. 
The extremely narrow band also supports temporally localized modes in each modulation period, where the associated Floquet mode-locking mechanism points to potential applications in generating mode-locked pulses. 
When the temporal modulation pattern follows a quasiperiodic sequence, periodicity in the time domain is replaced by long-range temporal order controlled by a phason parameter \cite{ni2025topological}.
The additional phason parameter can be regarded as the momentum reciprocal to a synthetic space, which spans a two-dimensional synthetic parameter space and hence allows the definition of Chern number, topological edge states, and a temporal analogue of Thouless pumping [see Fig. \ref{figure.9}(b)]. 
More generally, by applying the two-symbol substitutional sequence framework, one can study the band structure and wave-transport properties of aperiodic PTCs, including Thue-Morse and Fibonacci orders, which may exhibit multiscale spectral response and anomalous localization \cite{coppolaro2025exploring}. 
For the case of disordered time-varying media, it has been noted that the growth of wave energy initially follows a negative exponential distribution, while it approaches a log-normal distribution after a sufficiently long time \cite{carminati2021universal}. 
Another study on disordered PTC shows that temporal randomness can exponentially slow and eventually stop a propagating pulse while its intensity continues to increase \cite{sharabi2021disordered}. 
Experiments in water waves with randomly shifted temporal interfaces confirm the disorder-dependent exponential growth of the wave field, and further reveal that the disorder modifies the phase coherence associated with the original momentum gap \cite{apffel2022experimental}. 
The growth of modes in the momentum gap is suppressed, while the modes outside the momentum gap can be exponentially amplified [see Fig. \ref{figure.9}(c)]. 
In an effective temporal-lattice description, temporal disorder can lead to an analog of Anderson localization in the time domain, where an electromagnetic field can reach its maximum at a certain moment in time, and then decay exponentially \cite{eswaran2025anderson}. 
The relative phases between different modulation channels can also offer a degree of freedom for temporal-structure engineering, which can be employed as a symmetry-breaking tool and hence allows nonreciprocal negative refraction \cite{tavakol2026nonreciprocal}. 

\begin{figure}[!t]
    \centering
    \includegraphics[width=17cm]{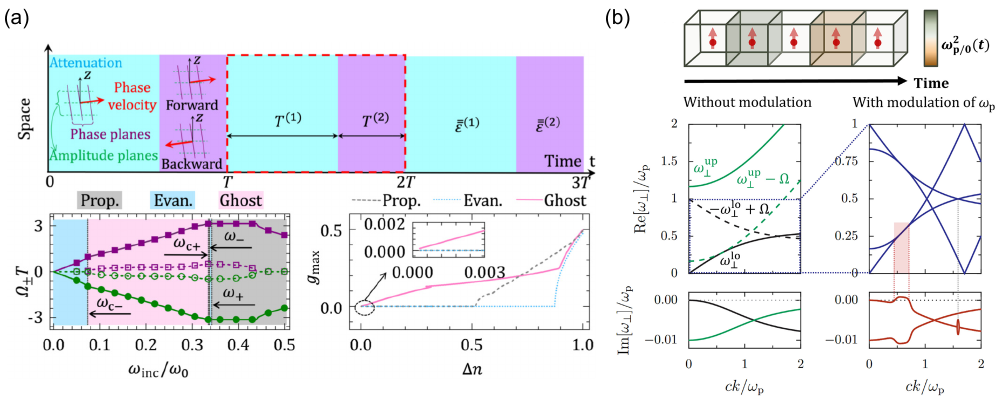}
    \caption{(a) Schematic of a biaxial anisotropic PTC, corresponding band structure versus the incident frequency $\omega_{\mathrm{inc}}$, and maximum amplification strength $g_{\mathrm{max}}$ of momentum band gaps for propagating, evanescent, and ghost waves versus modulation depth $\Delta n$ \captioncite{dong2025nonuniform}. 
    (b) Schematics of a dispersive PTC, where the plasma or resonance frequency of a Drude-Lorentz medium is modulated in time with an oscillating dipole inside. 
    Complex band structure of the static medium and dispersive PTC with modulated plasma frequency, where the dashed lines denote the Floquet replicas \captioncite{allard2026broadband}.
    Panel (a) is reproduced with permission
from Ref.~\captioncite{dong2025nonuniform},
\textcopyright~2025 American Physical Society;
panel (b) is reproduced from
Ref.~\captioncite{allard2026broadband},
published by the American Physical Society
under the Creative Commons Attribution
4.0 International License (CC BY 4.0).}\label{figure.10}
\end{figure}

Another degree of freedom lies in the electromagnetic medium itself, allowing scalar and nondispersive materials to be replaced by anisotropic, dispersive, or multimode systems.
For example, periodically switching between isotropic and biaxially anisotropic media extends momentum-gap physics from propagating waves to evanescent and ghost waves \cite{dong2025nonuniform}. 
In particular, ghost waves can support broad momentum gaps even for small modulation depths [see Fig. \ref{figure.10}(a)]. 
Furthermore, their gap modes exhibit an unusual imbalance between time-refraction and time-reflection channels, enabling selective amplification or attenuation of nonuniform fields.  
Additionaly, anisotropy also suggests an anisotropic band structure, enabling directional control of stationary charge radiation \cite{li2023stationary}. 
The inclusion of bianisotropic responses can introduce material-induced non-Hermiticity, which allows broadband temporal localization and delocalized temporal edge states \cite{jiang2026broadband}.
Another important aspect of material response is its frequency dispersion, which provides an additional degree of freedom for engineering PTCs. The dispersionless model introduced in Section 3.3 assumes an instantaneous constitutive relation, $D(t)=\varepsilon(t)E(t)$. 
In a dispersive time-varying medium, however, the electric displacement depends on the electric field at earlier times and can be expressed as $D(t)=\int_{-\infty}^{t}\varepsilon(t,t')E(t') dt'$, where $\varepsilon(t,t')$ depends on both the observation time $t$ and the excitation time $t'$, rather than only their difference as in a time-invariant medium.
 A practical approach is to introduce auxiliary dynamical variables, such as the polarization and polarization current in a time-dependent Drude-Lorentz medium. Together with the electromagnetic fields, these variables form an enlarged first-order evolution equation, which can be treated using the Floquet matrix framework introduced in Section 3.1. Material damping can also be incorporated into this formulation \cite{asgari2024theory,sun2025formulation}.
Beyond extending the theoretical description, material dispersion can substantially reshape the Floquet spectrum and modify the location, width, and amplification characteristics of momentum gaps. For example, modulating the plasma or resonance frequency of a dispersive medium modifies the coupling between the material's eigenmodes and their Floquet replicas [see Fig. \ref{figure.10}(b)]. Such effects can enable broadband dipole absorption in dispersive and absorptive PTCs without relying on the conventional narrow parametric-resonance window or gap-edge exceptional points \cite{allard2026broadband}. Dispersion engineering can also relax the required modulation speed and strength. In particular, plasma-frequency modulation combined with spatial nonlocality can produce momentum gaps of infinite extent \cite{salehi2026nonlocal}, while non-Foster materials can support momentum gaps extending down to zero momentum, with a gain rate that increases linearly with momentum \cite{li2025non}.
When the underlying platform supports several guided modes, temporal modulation can couple not only opposite-frequency branches of the same mode, but also different modal branches. 
Such intermodal coupling produces tilted momentum gaps whose centers are not fixed at half the modulation frequency \cite{li2026multi}. 

Beyond engineering modulation pattern or material response, one can also introduce additional physical mechanisms, such as nonlinearity, in PTCs and explore their novel phenomena. 
\begin{figure}[!b]
    \centering
    \includegraphics[width=17cm]{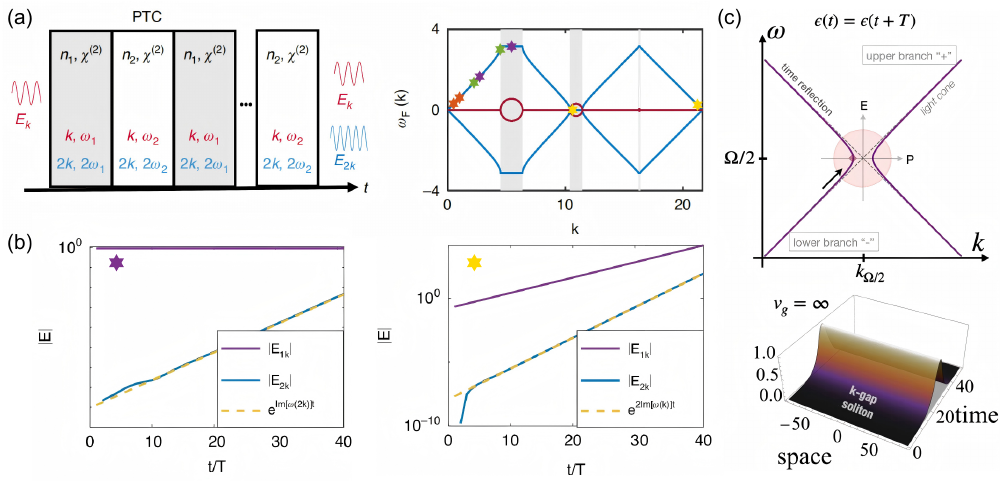}
    \caption{(a) Schematics of a nonlinear PTC and the band structure of linear PTC with fundamental second harmonic signal satisfying different relations, as denoted by the colored stars \captioncite{konforty2025second}. 
    (b) Phase-mismatched SHG with the second harmonic mode inside the momentum gap (purple star) and the fundamental mode (yellow star) inside the momentum gap, respectively \captioncite{konforty2025second}. 
    (c) Illustration of the $k$-gap soliton at the band edge \captioncite{pan2023superluminal}.
    Panels (a) and (b) are reproduced from
Ref.~\captioncite{konforty2025second},
\textcopyright~The Author(s) 2025 (CC BY 4.0);
panel (c) is reproduced with permission
from Ref.~\captioncite{pan2023superluminal},
\textcopyright~2023 American Physical Society.}\label{figure.11}
\end{figure}
For example, in a second-order nonlinear PTC [see Fig. \ref{figure.11}(a)], the exponentially growing momentum-gap modes can significantly enhance second-harmonic generation even when the conventional phase-matching condition is not satisfied, as illustrated in Fig. \ref{figure.11}(b) \cite{konforty2025second}. 
The enhancement of frequency harmonic generation can even extend to higher orders, resulting in cascaded exponentially amplification. 
On the other hand, a temporal soliton arises from the interplay between Floquet instability and Kerr nonlinearity \cite{pan2023superluminal}. 
The light field with its $k$ component at the band edge initially experiences exponential growth until it reaches a peak, where the Kerr nonlinearity shifts the band structure due to the increasing light field, and eventually brings the $k$ component outside the momentum gap, followed by a decaying intensity. 
Such interplay generates a periodically self-reconstructing soliton profile in the time domain, while it takes a plane-wave-like spatial profile, as shown in Fig. \ref{figure.11}(c). 
By further seeding a spatially local perturbation onto the temporal soliton, the nonlinear saturation is broken, which nucleates a breathing $k$-gap event \cite{zhang2026breathing}. 

\subsection{Synthetic frequency dimensions: a different view on time-periodic systems}

The Floquet formulation not only provides a method for calculating the quasifrequency spectrum, but also gives an alternative lattice picture of a time-periodic system. 
Recalling the eigenvalue equation in the enlarged Hilbert space in Eq. (\ref{eq36}), we now regard each harmonic index $m$ as a site along a synthetic frequency axis. 
Introducing a synthetic basis $\{\left|m\right>\}$, the Floquet Hamiltonian takes the lattice form
\begin{equation}
\mathcal{H}_{\mathrm{F}}=\sum_m (H_0-m\Omega) \otimes \left|m\right>\left<m\right| + \sum_{n \neq m} H_{m-n} \otimes \left|m\right>\left<n\right|.
\label{eq52}
\end{equation}
In this sense, the static part $H_0$ determines the internal structure of each frequency site, and the term $m\Omega$ denotes the frequency offset. 
The first-harmonic modulation couples the $m$-th site to $m\pm1$-th site, analogous to nearest-neighbor hopping, whereas higher modulation harmonics generate long-range couplings. 
The Floquet and frequency-lattice pictures thus describe the same time-periodic problem but emphasize different aspects of its physics. 
The Floquet picture focuses on the quasifrequency spectrum and the evolution over one modulation period, while the synthetic-frequency picture resolves how a Floquet state is distributed among its harmonic components and how energy is transported between them. 

\begin{figure}[!t]
    \centering
    \includegraphics[width=8cm]{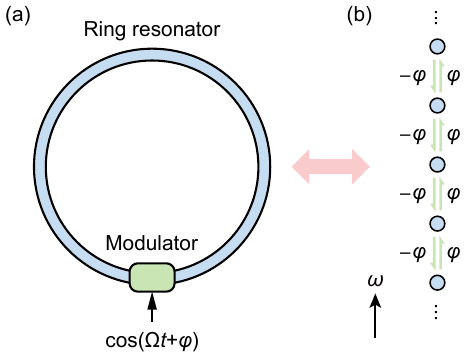}
    \caption{A dynamically modulated ring resonator in (a) can be mapped to a one-dimensional lattice in the synthetic frequency dimension in (b).}\label{figure.12}
\end{figure}

To illustrate how analogous frequency-space couplings can be engineered in a physical platform, we consider a dynamically modulated ring resonator in Fig.~\ref{figure.12}(a), where discrete cavity modes serve as synthetic lattice sites and temporal modulation induces coupling between them \cite{yuan2016photonic,ozawa2016synthetic,yuan2021synthetic}. 
Consider a ring resonator supporting a set of resonant frequencies $\omega_m=\omega_0+m\Omega_{\mathrm{R}}$, where $\Omega_{\mathrm{R}}=2\pi v_g /L$ is the free spectral range, $v_g$ is the intracavity group velocity, and $L$ is the length of the ring. 
A phase modulator is then placed in the cavity, which produces a periodic permittivity variation in the modulated region
\begin{equation}
\varepsilon(t)=\bar{\varepsilon}+\delta \varepsilon \cos (\Omega_{\mathrm{M}} t+\phi_\mathrm{M}),
\label{eq53}
\end{equation}
with a modulation frequency close to the free spectral range $\Omega_{\mathrm{M}}-\Omega_{\mathrm{R}}=\Delta$, which produces a dynamically varying phase shift on the light field
\begin{equation}
\mathcal{E}(t^+)=\mathcal{E}(t^-)e^{i\alpha \cos (\mathrm{\Omega_{\mathrm{M}}}t+\phi_\mathrm{M})}.
\label{eq54}
\end{equation}
Here, $\mathcal{E}(t^{\pm})$ is the modal amplitude of the electric field before (after) passing through the modulator. 
Such a modulation generates sidebands separated by $\Omega_{\mathrm{M}}$. 
For a small frequency detuning $|\Delta|\ll\Omega_{\mathrm R}$, these sidebands overlap resonantly with adjacent cavity modes, enabling transitions between $\omega_m$ and $\omega_{m\pm 1}$. 
Under weak modulation, where the higher-order sidebands are neglected, the modal dynamics reduces to a nearest-neighbor tight-binding Hamiltonian
\begin{equation}
H_{\mathrm{syn}}=g\sum_m\left[ e^{-i (\mathrm{\Delta}t+\phi_{\mathrm{M}})} a^{\dagger}_m a_{m+1}+e^{i (\mathrm{\Delta}t+\phi_{\mathrm{M}})}a^{\dagger}_{m+1 }a_m\right],
\label{eq55}
\end{equation}
where $a_{m}$ ($a^{\dagger}_{m}$) is the annihilation (creation) operator of the $m$-th resonant mode, and $g=-\alpha \Omega_{\mathrm{R}}/4\pi$. 
By further applying the transformation $a_m=e^{i\Delta m t} b_m$, Eq. (\ref{eq55}) is rewritten as a time-independent Hamiltonian \cite{yuan2016bloch}
\begin{equation}
\tilde{H}_{\mathrm{syn}}= \sum_m  m \Delta b^{\dagger}_m b_m+g\sum_m\left[ e^{-i \phi_{\mathrm{M}}} b^{\dagger}_m b_{m+1}+e^{i \phi_{\mathrm{M}}}b^{\dagger}_{m+1 } b_m\right].
\label{eq56}
\end{equation}
Therefore, the discrete resonant frequencies act as lattice sites, while modulation-induced frequency conversion connects the neighboring sites, as depicted in Fig. \ref{figure.12}(b), which allows the modulation parameters to have direct physical meaning in the synthetic model. 
The modulation amplitude $\alpha$ determines the hopping strengths, and the modulation $\phi_\mathrm{M}$ is encoded in the hopping phase between the lattice sites. 
The detuning $\Delta$ acts as a linear potential along the synthetic frequency dimension, acting as a synthetic force on a wave packet in the synthetic frequency space and hence leading to spectral Bloch oscillations \cite{yuan2016bloch}. 
Furthermore, the model is not restricted to nearest-neighbor coupling. A modulation component with frequency $q \Omega_{\mathrm{R}}$ can connect the $m$-th and $(m+q)$-th modes, giving rise to long-range hopping. 

A direct consequence of the modulation-induced hopping is that the time-modulated ring resonator can serve as a platform for exploring artificial gauge fields and topological photonics. 
In particular, the hopping phase determined by the modulation phase $\phi_\mathrm{M}$ mimics a Peierls phase induced by an effective lattice gauge field $\bm{\mathcal{A_{\mathrm{eff}}}}$ along the frequency axis \cite{peierls1933theorie}
\begin{equation}
\phi_\mathrm{M}=\int^{m+1}_{m} \bm{\mathcal{A_{\mathrm{eff}}}} \cdot   d \bm{l}.
\label{eq57}
\end{equation}
A uniform hopping phase in a one-dimensional chain can generally be removed by a gauge transformation and is therefore not observable by itself. 
Its physical significance emerges when the frequency axis is combined with another degree of freedom to form a gauge-invariant synthetic magnetic flux, such as resonator index \cite{yuan2016photonic,ozawa2016synthetic}, pseudospins \cite{dutt2020single}, or additional long-range hoppings \cite{yuan2018synthetic, cheng2023multi}. 
	By carefully designing the modulation and the system, various topological phases associated with Abelian \cite{yuan2019photonic, dutt2020higher, li2023direct, suh2024photonic, pellerin2024wave, ye2025construction} or non-Abelian gauge fields \cite{cheng2023artificial, pang2024synthetic, cheng2025non, dong2025sl} can be simulated. 
For a comprehensive review on recent progress in synthetic frequency dimensions, please refer to Ref. \cite{yu2025comprehensive}. 

This frequency-domain perspective also manifests in the experimental synthetic-dimension platforms discussed in Section 5.3, where couplings between frequency modes can be engineered through various modulation schemes. 
Nevertheless, the correspondence between such effective lattice models and time-varying Maxwell systems requires careful distinction. 
In particular, switching an effective Hamiltonian in synthetic space does not necessarily correspond to an abrupt change in the electromagnetic constitutive parameters, and therefore need not impose the same temporal interface conditions. 
Moreover, the quasimomentum defined in a synthetic lattice should not be confused with the physical wavevector conserved at a temporal interface occurring in a spatially uniform medium.

\section{Space-time systems}

\subsection{Space-time interfaces}

\begin{figure}[!t]
    \centering
    \includegraphics[width=16cm]{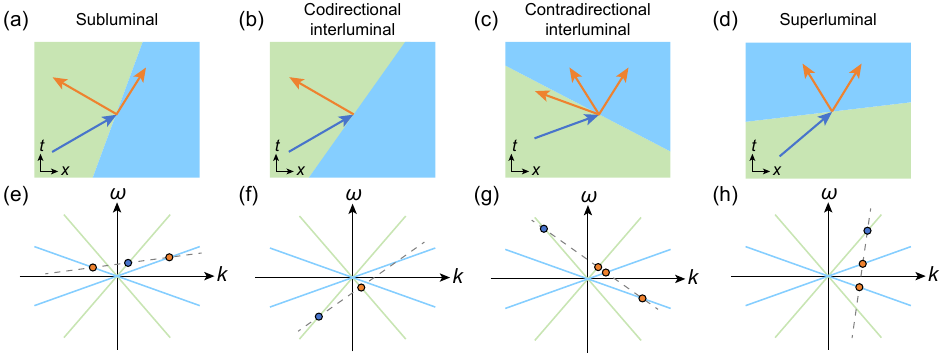}
    \caption{Illustration of the scattering of light at a (a) subluminal, (b) codirectional interluminal, (c) contradirectional interluminal, and (d) superluminal interface. 
    Corresponding projections of the incident wave at a (e) subluminal, (f) codirectional interluminal, (g) contradirectional interluminal, and (h) superluminal interface.}\label{figure.13}
\end{figure}

The systems discussed in the preceding sections are purely varying in the time domain, while their spatial structure remains homogeneous. 
In this section, we consider the systems modulated in both space and time, namely, the space-time systems. 
We start with the simplest cases, i.e., a space-time interface. 
Recall that a spatial interface preserves time-translation symmetry and therefore conserves frequency, whereas a temporal interface preserves spatial translation symmetry and conserves momentum. 
A space-time interface breaks both symmetries separately, allowing the frequency and momentum of a wave to change simultaneously. 
Nevertheless, when the interface moves uniformly, a spatiotemporal translation symmetry remains along its trajectory, yielding a generalized conservation relation that constrains the scattering process. 
A general space-time interface may be represented by $f(z,t)$, and the variation of the medium can be written as \cite{plansinis2015temporal, deck2019uniform}
\begin{equation}
n(z,t)=n_0+\Delta n \Theta [f(z,t)],
\label{eq58}
\end{equation}
where $\Theta(\cdot)$ is the Heaviside function. 
For a uniformly moving interface, $f(z,t)$ is given by
\begin{equation}
f(z,t)=z-v_{\mathrm{m}}t.
\label{eq59}
\end{equation}
Here, $v_{\mathrm{m}}$ denotes the velocity of a traveling modulation, such as a pump-induced refractive-index front, instead of the mechanical motion of matter. 
Therefore, such a modulation can access luminal and superluminal velocities without transporting matter or information faster than light. 
In the limiting cases with $v_{\mathrm{m}}=0$ and $v_{\mathrm{m}}\rightarrow \infty$, the space-time interface reduces to a static spatial interface and a spatially uniform temporal interface, respectively. 
We label the wave velocities on two sides of the space-time interface as $v_1$ and $v_2$ (assuming $v_1>v_2$), respectively.  
The space-time interface can then be classified into three regimes according to the relation between the modulation velocity $v_{\mathrm{m}}$ and the wave velocities $v_1$ and $v_2$ \cite{deck2019uniform}
\begin{equation}
\begin{aligned}
|v_{\mathrm{m}}| < v_2 & \rightarrow \mathrm{subluminal ~interface},\\
v_2 < |v_{\mathrm{m}}| & < v_1 \rightarrow \mathrm{interluminal~ interface},\\
|v_{\mathrm{m}}| > v_1 & \rightarrow \mathrm{superluminal ~interface}.
\end{aligned}
\end{equation}
A subluminal interface behaves like a moving spatial interface, where an incident wave is partially reflected and transmitted, as illustrated in Fig. \ref{figure.13}(a). 
The interluminal regime exhibits distinct scattering behavior as the interface moves faster than waves in one medium but slower than the other. 
For codirectional incidence, the incident wave can reach the interface, but the transmitted wave cannot propagate ahead of it, leaving only a reflected wave [Fig. \ref{figure.13}(b)]. 
In contrast, for contradirectional incidence, the interface moves toward the incident wave and can generate three outgoing waves [Fig. \ref{figure.13}(c)] \cite{deck2019uniform,deck2019scattering, yang2026twin}. 
A superluminal interface, on the other hand, sweeps across the incident field and more closely resembles a temporal interface, generating forward- and backward-propagating waves [Fig. \ref{figure.13}(d)]. 
In the luminal cases, where the interface velocity matches the wave velocity in one of the media $v_{\mathrm{m}}=v_i$, the interface coincides with the wave trajectory, which may lead to wave accumulation and shock-wave formation. 

Although independent spatial and temporal translation symmetries are broken, a uniform modulation of the form $n(z,t)=n(z-v_{\mathrm{m}}t)$ remains invariant under the combined transformation $z \rightarrow z+ \Delta z$ and $t \rightarrow t+ \Delta z/v_{\mathrm{m}}$, which ensures a spatiotemporal continuous symmetry. 
By Noether's theorem, such a spatiotemporal continuous symmetry give rise to the energy-momentum conservation \cite{liberal2024spatiotemporal}
\begin{equation}
\omega-v_{\mathrm{m}}k=\mathrm{constant} \Longleftrightarrow \Delta \omega =v_{\mathrm{m}}\Delta k.
\label{eq61}
\end{equation}
Thus, energy and momentum are not conserved separately, but instead, their changes are linked by the modulation velocity. 
In the spatial limit $(v_{\mathrm{m}}=0)$, this relation reduces to energy conservation, whereas in the temporal limit  $(v_{\mathrm{m}}\rightarrow \infty)$ it reduces to momentum conservation.
The allowed scattering channels can be identified geometrically in the dispersion diagrams. 
By drawing an oblique line through the point representing the incident wave $(\omega_\mathrm{i}, k_{\mathrm{i}})$, the frequencies and momenta of the reflected and transmitted waves are found at the intersections with the dispersion curves on either side of the interface [see Figs. \ref{figure.13}(e)-(h)]. 
Notably, these intersections must also satisfy the causal requirement, i.e., the scattered waves must propagate away from the moving interface. 
In the interluminal regime, this leads to only one outgoing reflected wave for codirectional incidence [Fig. \ref{figure.13}(f)], but three outgoing waves for contradirectional incidence [Fig. \ref{figure.13}(g)]. 
Hence, varying the interface velocity changes not only the frequencies and momenta of the scattered waves, but also the number of physically accessible scattering channels. 
In dispersive media, the absence of a propagating transmitted intersection can produce a temporal analogue of total internal reflection \cite{plansinis2015temporal}. 
Two parallel moving interfaces satisfying the temporal total-internal-reflection condition can confine a pulse within a moving temporal window, forming a temporal waveguide with discrete guided modes \cite{plansinis2016temporal}. 

To determine the amplitudes of the scattered waves, we then consider the interface conditions at a space-time interface. 
These conditions may be derived by transforming to a frame where the interface becomes spatial or temporal, applying the familiar continuity relations, and transforming back to the laboratory frame \cite{deck2019uniform}. 
For a one-dimensional interface moving along $z$, with transverse components $E_x$ and $H_y$, the generalized continuity relations may be written as
\begin{equation}
    \begin{aligned}
    E_{x,1}-v_{\mathrm{m}}B_{y,1}&=E_{x,2}-v_{\mathrm{m}}B_{y,2},\\
    H_{y,1}-v_{\mathrm{m}}D_{x,1}&=H_{y,2}-v_{\mathrm{m}}D_{x,2}.
    \end{aligned}
    \label{eq62}
\end{equation}
These conditions can return to the familiar limits: $E_x$ and $H_y$ become continuous when $v_{\mathrm{m}}=0$, and $B_y$ and $D_x$ become continuous when $v_{\mathrm{m}} \to \infty$. 
The amplitudes of the transmitted and reflected waves can be obtained from these interface conditions using the wave impedance $Z_i=\sqrt{\mu_i/\varepsilon_i}$, which relates the electric and magnetic field amplitudes in each medium through $H_y^{\pm}=\pm E_x^{\pm}/Z_i$. Consequently, the scattering coefficients depend on the impedance mismatch between the two media as well as the interface velocity. In the stationary limit, the electric-field amplitude $R$ reduces to the familiar Fresnel expression $R=(Z_2-Z_1)/(Z_2+Z_1)$ at normal incidence.
Note that the interluminal cases with multiple outgoing channels may require additional constraints. 
For a nonuniformly moving interface, the velocity varies along its trajectory and no single global quantity remains conserved, yet this additional freedom can enable more general frequency transformations \cite{wu2024conformal}.

\begin{figure}[!t]
    \centering
    \includegraphics[width=17cm]{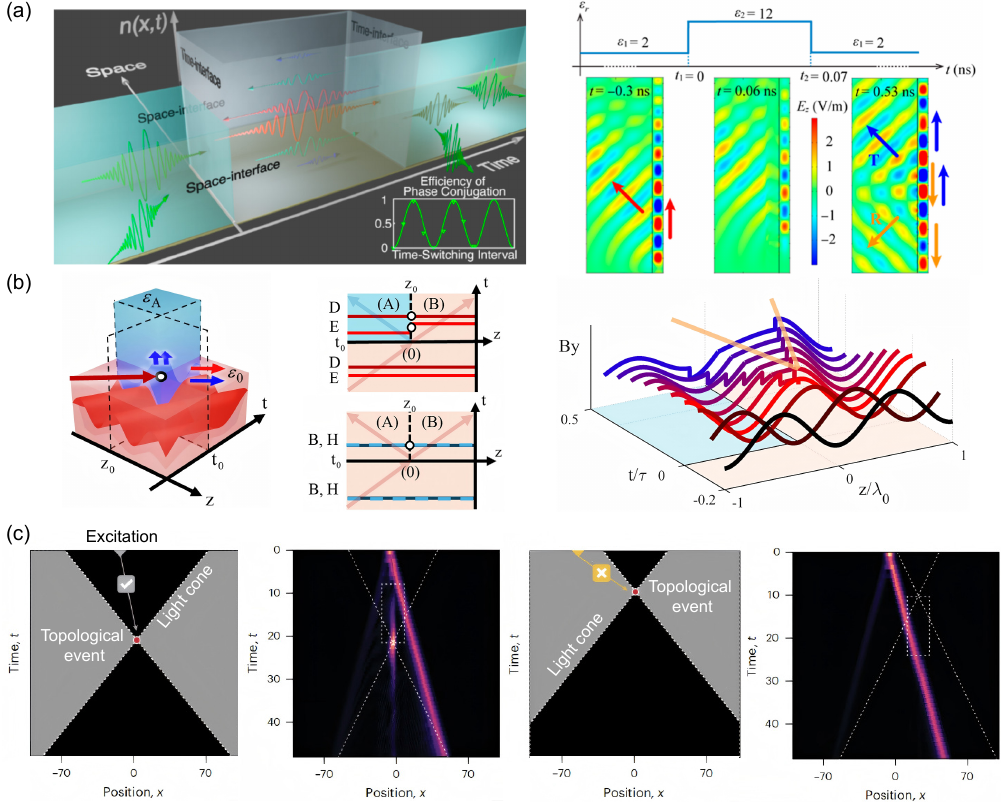}
    \caption{(a) Illustration of a space-time leaky waveguide, time variation of the permittivity in the waveguide region, and simulated electric field snapshots before, between, and after the two temporal switching events, where $T$ denotes the forward wave and $R$ denotes the phase-conjugated wave \captioncite{yin2022efficient}. 
    (b) Illustration of a space-time corner, distribution of electric field and magnetic field, and space-time evolution of a monochromatic wave impinging on a space-time corner \captioncite{stefanini2025theory}. 
    (c) Coupling of a space-time topological event, which is only available from its past light cone, while it is forbidden from its future light cone due to causality \captioncite{feis2025space}. 
    Panel (a) is reproduced from Ref.~\captioncite{yin2022efficient}, \textcopyright~2022 American Chemical Society (CC BY-NC-ND 4.0);
panel (b) is reproduced with permission from Ref.~\captioncite{stefanini2025theory}, \textcopyright~2025 American Physical Society;
panel (c) is reproduced from Ref.~\captioncite{feis2025space}, \textcopyright~The Author(s) 2025 (CC BY 4.0).}\label{figure.14}
\end{figure}

More generally, space-time scattering structures can be formed by combining spatial and temporal interfaces rather than by moving a single material boundary. 
A representative case is a spatially bounded region that is switched between different states over a finite time interval. 
In a resonant structure, this combination of spatial and temporal interfaces substantially enhances the interaction with the modulation. 
For example, a finite leaky-wave cavity in Fig. \ref{figure.14}(a) can transfer the field between positive- and negative-frequency modes by two temporal interfaces, and thereby realize efficient phase conjugation without switching the surrounding space. 
Another situation is a space-time corner, where a spatial interface and a temporal interface intersect at a single point in space-time as shown in Fig. \ref{figure.14}(b). 
This enables wave conversion between frequency and momentum, or in other words swapping between positive- and negative-index modes \cite{yin2022efficient}. 
In this case, the usual spatial continuity relations for $\bm{E}$ and $\bm{H}$ and the temporal continuity relations for $\bm{D}$ and $\bm{B}$ cannot always be imposed independently at this intersection \cite{stefanini2025theory}. 
Their incompatibility produces an additional localized impulsive response, which generates shock waves. 
The corner must therefore be treated as an additional scattering object rather than as a simple sequence of separate spatial and temporal events. 
Furthermore, the combination of a spatial and temporal topological domain wall can confine a wave around their intersection, producing a localized event in space-time \cite{feis2025space}. 
However, we should note that only excitations lying within the past light cone of this space-time point can couple into it due to causality, as illustrated in Fig. \ref{figure.14}(c), reflecting the fact that time dimension is not equivalent to an additional spatial coordinate.

\subsection{Traveling-wave space-time crystals}

With the understanding of space-time interfaces, we can now move to a space-time crystal, which is a medium that is periodically modulated in both space and time. 
We first consider the simplest case with traveling-wave modulation and give the basic theoretical framework for analyzing such a system. 
Consider a Hamiltonian in the form \cite{gao2021floquet-bloch, peng2022topological, zhang2025topolectrical}
\begin{equation}
H(x,t)=H(G x-\Omega t),
\label{eq63}
\end{equation}
with a spatiotemporal periodicity $H(G x-\Omega t)=H(G (x+a)-\Omega (t+T))$, where $a=2\pi/G$, $T=2\pi/\Omega$, and the velocity of the modulation is $v_{\mathrm{m}}=\Omega/G$. 
Unlike a system with independent spatial and temporal periodicities, a traveling-wave space-time crystal is characterized by one space-time reciprocal vector $(G, \Omega)$, corresponding to an oblique energy-momentum Brillouin zone. 
We can therefore expand the Hamiltonian and the state as
\begin{equation}
\begin{aligned}
H(x,t)&=\sum_m H_m e^{i m (Gx-\Omega t)},\\
|\psi(x,t)\rangle &=e^{i(k x-\omega t)}\sum_m \left|u_m \right> e^{i m(Gx-\Omega t)},
\end{aligned}
\label{eq64}
\end{equation}
indicating that the $m$-th Bloch-Floquet harmonic carries both a shifted momentum $k_m=k+m G$ and frequency $\omega_m=\omega+m \Omega$. 
Thus, the traveling-wave modulation couples states separated by multiples of the space-time reciprocal vector $(G, \Omega)$, giving rise to band folding along an oblique direction in the $(k, \omega)$ space. 
By substituting Eq. (\ref{eq64}) into the Schr\"odinger-type equation, one can obtain a generalized eigenvalue problem in the enlarged Hilbert space
\begin{equation}
\sum_m \left[ H_{n-m}(k+mG)-m\Omega \delta_{nm} \right] \left| u_m \right> = \omega \left| u_n \right>.
\label{eq65}
\end{equation}
The diagonal blocks represent the uncoupled Floquet-Bloch replicas, while the off-diagonal blocks $H_{n-m}$ encode the couplings between different replicas. This formulation reduces to conventional Floquet theory for $G=0$, and spatail Bloch extension for $\Omega=0$. 

For a piecewise-constant traveling-wave profile, the harmonic-space Hamiltonian can be replaced by a transfer-matrix construction \cite{deck2019uniform}. 
Each discontinuity is described by a moving-interface matrix $T_{ji}$ obtained from the generalized interface conditions in Eq. (\ref{eq62}), while the propagation through a homogeneous space-time segment is represented by a diagonal matrix $P_j$. The transfer matrix of one bilayer unit cell is schematically
\begin{equation}
M_{\mathrm{ST}}=P_2 T_{21}P_1 T_{12},
\label{eq66}
\end{equation}
which is the generalization of the conventional transfer matrix method in the spatial and temporal cases. 
Bloch-Floquet periodicity requires
\begin{equation}
M_{\mathrm{ST}}\left|\psi\right>=e^{i\Phi_{\mathrm{ST}}}\left|\psi\right>,
\label{eq67}
\end{equation}
where $\Phi_{\mathrm{ST}}$ is the Bloch-Floquet phase. 
The solution to this eigenproblem can be found by substituting Eq. (\ref{eq66}) into Eq. (\ref{eq67}), which gives the dispersion relation. 
Alternatively, spatial and temporal unit-cell matrices may be constructed separately in the laboratory frame, providing an orthogonal Bloch-Floquet formulation that highlights space-time duality \cite{deck2023orthogonal}. 
For finite crystals in the space-time domain, the scattering response also depends on the orientation of the truncating boundaries in the space-time plane, since different truncations impose different phase-matching conditions with the surrounding medium \cite{deck2019uniform}. 

\begin{figure}[!t]
    \centering
    \includegraphics[width=16cm]{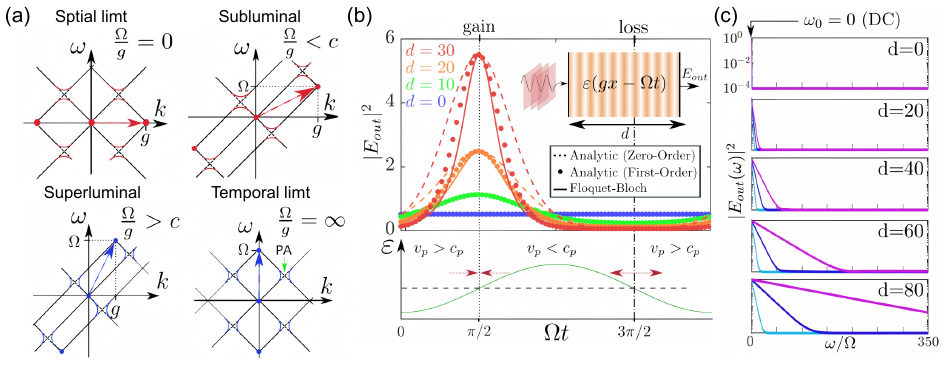}
    \caption{(a) Band structure of a spatial crystal, subluminal crystal, superluminal crystal, and temporal crystal \captioncite{galiffi2019broadband}. 
    (b) An incident plane wave is compressed and exponentially amplified as it propagates through a luminal metamaterial. At the points where the velocity of the waves matches the velocity of modulation, the gain becomes maximum, and these points act as attractors \captioncite{galiffi2019broadband}. 
    (c) Power spectra of the transmitted waves for a DC input \captioncite{galiffi2019broadband}. 
    Panels (a)-(c) are reproduced with permission from Ref.~\captioncite{galiffi2019broadband}, \textcopyright~2019 American Physical Society.}\label{figure.15}
\end{figure}

The band structure of a traveling-wave space-time crystal can also be classified according to the relation between the modulation velocity and characteristic wave velocity, as illustrated in Fig. \ref{figure.15}(a) \cite{galiffi2019broadband}. 
In the subluminal regime, the space-time reciprocal vector is predominantly space-like, and Bragg coupling generally opens frequency gaps associated with spatial attenuation which reduces to the spatial limit for $v_{\mathrm{m}}=0$. 
On the other hand, in the superluminal regime, it becomes predominantly time-like, producing momentum gaps in which conjugate complex quasifrequencies describe temporal growth and decay, reducing to temporal limit for $v_{\mathrm{m}} \rightarrow \infty$. 
The transition between the subluminal and superluminal regimes occurs where the modulation velocity approaches a local wave velocity, i.e., the luminal regime. 
In this scenario, the forward propagating Bloch-Floquet harmonics become nearly degenerate and strongly interact, where conventional Floquet-Bloch description in terms of extended Bloch modes and a well-defined dispersion relation breaks down \cite{galiffi2021photon}. 
As illustrated in Fig. \ref{figure.15}(b), waves approaching the gain point $\Omega t=\pi/2$ from ahead (behind) encounter a lower (higher) permittivity and therefore propagate faster (slower), being attracted toward the gain point \cite{galiffi2019broadband}. 
Consequently, the field is driven toward the gain point and exponentially amplified, resulting in the generation of a pulse train. 
By contrast, the corresponding velocity profile near the loss point $\Omega t=3\pi/2$ drives waves away from it, resulting in local energy depletion. 
Such a mechanism enables broadband nonreciprocal gain and the generation of supercontinuum [see Fig. \ref{figure.15}(c)], which does not rely on backscattering in contrast to conventional parametric amplification. 
A complementary physical interpretation attributes the gain to the compression of conserved electric and magnetic lines of force \cite{pendry2021gain1,pendry2021gain}. 
In impedance-matched systems, the added energy may raise the photon frequencies without changing their number \cite{pendry2022photon}. 
Transitions across the negative-frequency branch allow spontaneous or stimulated photon-pair emission, connecting luminal localization to horizon-like radiation in transluminal gratings \cite{horsley2023quantum}.

Beyond the scalar real-permittivity traveling-wave modulation mentioned above, space-time modulation can act on complex constitutive parameters, anisotropic material tensors, or internal degrees of freedom such as polarization and angular momentum. These extensions introduce additional coupling channels and dynamical symmetries, allowing the effects of space-time modulation to extend beyond conventional frequency and momentum conversion. 
Independently controlling the real and imaginary components of a traveling-wave modulation introduces explicit non-Hermiticity, producing exceptional degeneracies, asymmetric momentum gaps, and harmonic-selective directional amplification \cite{moghaddaszadeh2022complex}. 
Traveling-wave-modulated media may also be embedded in a conventional spatial crystal, where the resulting nonreciprocity can generate a non-Hermitian skin effect \cite{ding2024non}. 
Moreover, traveling-wave modulation may act on the principal axes of anisotropic permittivity and permeability tensors rather than on scalar material parameters.
A propagating helical tensor pattern forms an optical Archimedes' screw, leading to polarization locking and chirality-selective parity-time-broken phases \cite{galiffi2022archimedes}. 
One may also replace linear translation by a combined temporal and rotational symmetry. 
Synthetic rotation of an anisotropic crystal preserves a combined space-time rotational symmetry, leading to conservation of a linear combination of energy and spin angular momentum (SAM) and producing Floquet sidebands with frequency-SAM locking \cite{liberal2026synthetic}. 
An angular traveling wave realizes Floquet rotation in a ring resonator, where the transition to the superluminal regime opens gaps in orbital angular momentum (OAM) and enables OAM-selective rotational superradiance \cite{nasari2026observation}. Furthermore, traveling-wave modulation can be implemented in discrete physical systems. The resulting space-time translational symmetry produces an oblique energy-momentum Brillouin zone and supports nontrivial $\pi$-gap edge states even in a monoatomic lattice \cite{tong2026acoustic}.

\begin{figure}[!t]
    \centering
    \includegraphics[width=17cm]{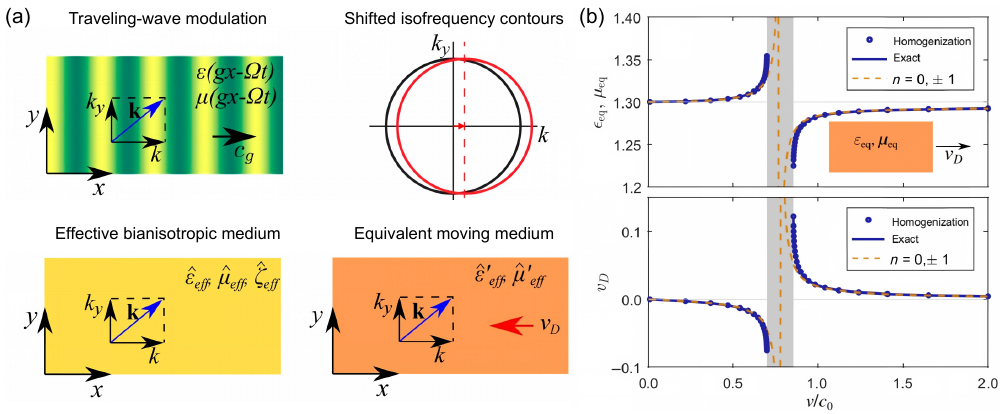}
    \caption{(a) A medium with space-time traveling-wave-modulated permittivity and permeability that features a shifted isofrequency contour. 
    In the long-wavelength limit, such a medium can be mapped to an effective bianisotropic medium, which is equivalent to a moving medium \captioncite{huidobro2019fresnel}. 
    (b) The effective permittivity, permeability and drag velocity versus the modulation velocity \captioncite{huidobro2021homogenization}. 
    Panel (a) is reproduced from Ref.~\captioncite{huidobro2019fresnel}, \textcopyright~The Author(s) 2019 (PNAS License to Publish);
panel (b) is reproduced with permission from Ref.~\captioncite{huidobro2021homogenization}, \textcopyright~2021 American Physical Society.}\label{figure.16}
\end{figure}

When the wavelength is much larger than the space-time period and the modulation velocity lies outside the luminal regime, the Bloch-Floquet formalism can be replaced by an effective-medium description. 
As illustrated in Fig. \ref{figure.16}(a), the traveling-wave modulations of both permittivity and permeability generate effective anisotropic permittivity and permeability together with a magnetoelectric coupling \cite{huidobro2019fresnel}. 
The resulting displacement of the isofrequency contours can be interpreted as a synthetic Fresnel drag produced by an equivalent moving medium, despite the absence of physical motion.
Notably, the drag velocity $v_{\mathrm D}$, together with the effective permittivity and permeability, reverses their sign across the luminal interval, allowing the drag direction to be reversed without changing the modulation direction [see Fig. \ref{figure.16}(b)]  \cite{huidobro2021homogenization}. 
On the other hand, a particular subclass of spacetime crystals, namely, the Minkowskian isorefractive spacetime crystals, can reproduce the electromagnetic response of a moving dielectric photonic crystal exactly, rather than only in the long-wavelength limit \cite{prudencio2023replicating}. 
Anisotropic unit cells and interlaced sublattices driven along different directions extend the long-wavelength effective response beyond the simple antisymmetric coupling of an isotropic moving medium, enabling a wide range of nonreciprocal bianisotropic couplings \cite{prudencio2024engineering}.

\subsection{Space-time crystals beyond traveling-wave modulation}

\begin{figure}[!t]
    \centering
    \includegraphics[width=12cm]{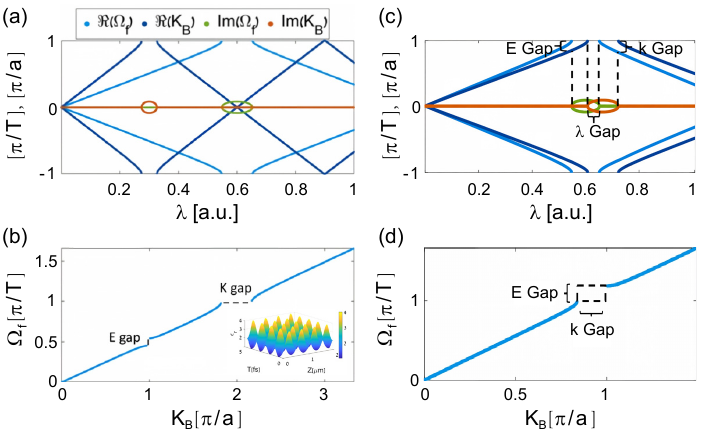}
    \caption{(a) Spatial and temporal bandstrucutre of a coordinate-separable space-time crystal, where the energy gap and momentum gap open separately \captioncite{sharabi2022spatiotemporal}. 
    (b) The unfolded band structure of the space-time crystal in (a). The inserted figure shows the spatiotemporal profile of the system \captioncite{sharabi2022spatiotemporal}. 
    (c) Spatial and temporal bandstrucutre of a coordinate-separable space-time crystal, where the energy gap and momentum gap overlap \captioncite{sharabi2022spatiotemporal}. 
    (d) The unfolded band structure of the space-time crystal in (c), which opens a mixed energy-momentum gap \captioncite{sharabi2022spatiotemporal}. 
    Panels (a)-(d) are reproduced from Ref.~\captioncite{sharabi2022spatiotemporal}, \textcopyright~2022 Optica Publishing Group (Optica Open Access Publishing Agreement).}\label{figure.17}
\end{figure}

Another type of space-time crystals can be constructed by independently prescribing the spatial structure and temporal drive. 
A typical example is a coordinate-separable space-time crystal \cite{sharabi2022spatiotemporal}
\begin{equation}
\varepsilon(z,t)=\varepsilon_0 \cdot \varepsilon_z(z) \cdot \varepsilon_t(t).
\label{eq68}
\end{equation}
This coordinate-separable form allows one to solve the eigenvalue problem of the system separately in space and time, where the spatial (temporal) part of the eigenvalue equation has the form of an exact photonic spatial (temporal) crystal with eigenvalues $K_{\mathrm{B}}$ ($\Omega_{\mathrm{f}}$). 
Therefore, the solution of the spatiotemporal system is written as the product of a Bloch mode and a Floquet mode
$E=\exp[i(K_{\mathrm{B}}z+ \Omega_{\mathrm{f}}t)]\cdot U_z(z)\cdot U_t(t)$, where $U_z(z)$ and $U_t(t)$ are periodic functions in their respective coordinates. 
Fig. \ref{figure.17}(a) shows the spatial Bloch spectrum $K_{\mathrm{B}}(\lambda)$ and temporal Floquet spectrum $\Omega_{\mathrm{f}}(\lambda)$, which are calculated with respect to the same dimensionless parameter $\lambda$ introduced by a coordinate-separable form of the Maxwell wave equation. 
In this case, the energy and momentum gaps occupy distinct ranges of $\lambda$. 
Pairing $K_{\mathrm{B}}(\lambda)$ and $\Omega_{\mathrm{f}}(\lambda)$ yields the unfolded spatiotemporal band structure in Fig. \ref{figure.17}(b), where the energy and momentum gaps remain separated.  
Varying the relative spatial and temporal periods shifts the two gaps and can bring them into overlap. 
As shown in Fig. \ref{figure.17}(c), the overlapping spatial and temporal spectra can contain three regions: an energy-gap region, a momentum-gap region, and a central region where both $K_{\mathrm{B}}$ and $\Omega_{\mathrm{f}}$ are complex, which results in a mixed energy-momentum gap in Fig. \ref{figure.17}(d). 
The mixed energy-momentum gap reflects the competition of two opposing effects. 
The imaginary part of $K_{\mathrm{B}}$ produces exponential attenuation in space, whereas the imaginary part of $\Omega_{\mathrm{f}}$ produces exponential growth or decay in time. 
Spatial localization dominates when $ |\mathrm{Im}(K_{\mathrm{B}})| a> |\mathrm{Im} (\Omega_{\mathrm{f}})| T$, while temporal amplification dominates when $ |\mathrm{Im}(K_{\mathrm{B}})| a< |\mathrm{Im} (\Omega_{\mathrm{f}})| T$.
Under certain conditions, $K_{\mathrm{B}}(\lambda)$ and $\Omega_{\mathrm{f}}(\lambda)$ can overlap completely, causing the projected energy and momentum gaps in the $(K_{\mathrm{B}}(\lambda), \Omega_{\mathrm{f}}(\lambda))$ plane to close and their associated exceptional points to coincide, which supports unique modes causing waves to act as a source, emitting radiation backwards.

Topological phenomena in spatiotemporal media can arise from both the temporal modulation of existing spatial topological states and the interplay between spatial and temporal periodicities. 
As an example of the former case, we consider periodic driving of a spatial topological interface, which generates Floquet replicas of the original edge state at frequencies $\Omega_0+m\Omega_{\mathrm{m}}$, while preserving its spatial localization. 
\begin{figure}[!b]
    \centering
    \includegraphics[width=17cm]{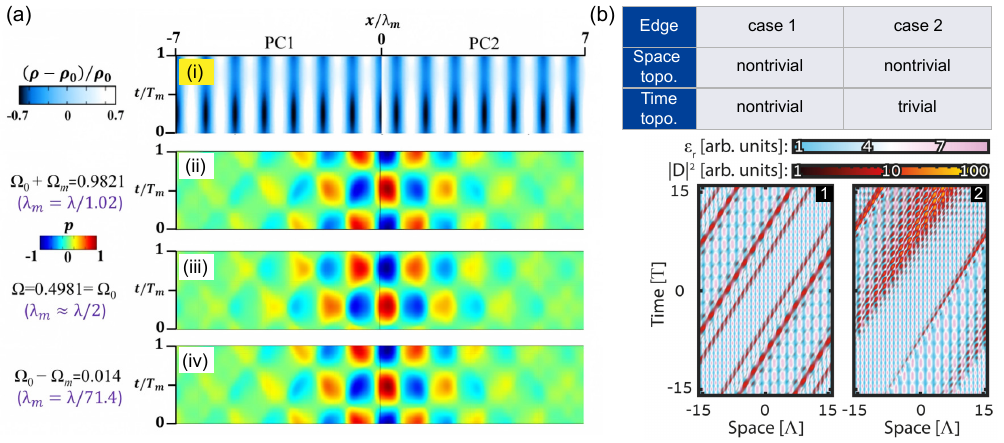}
    \caption{(a) Spatiotemporal profile of a topological interface formed between space-time crystals with different topologies, and the distribution of the topological edge states and their Floquet replicas \captioncite{oudich2019space}. 
    (b) Spatiotemporal evolution of the topological edge states for space-time interfaces with different combinations of spatial and temporal topologies \captioncite{segal2025two}. 
    Panel (a) is reproduced from Ref.~\captioncite{oudich2019space}, \textcopyright~2019 American Physical Society (CC BY 4.0);
panel (b) is reproduced with permission from Ref.~\captioncite{segal2025two}, \textcopyright~2025 American Physical Society.}\label{figure.18}
\end{figure}
This allows topological states associated with the Bragg regime to be excited at substantially lower frequencies, as illustrated in Fig. \ref{figure.18}(a) \cite{oudich2019space}. 
When spatial and temporal topologies coexist, their interplay can further determine the dynamics of states bound to an interface in space-time. 
For systems characterized by independent spatial and temporal topological indices, an interface across which both indices change can support a state localized transverse to the interface while propagating along it with real momentum and quasifrequency, without exponential amplification or decay. 
In contrast, changing only the spatial index can produce an interface state with complex quasifrequency and exponential amplification, as illustrated in Fig. \ref{figure.18}(b) \cite{segal2025two}. 
While these examples illustrate how temporal modulation modifies existing or combined spatial and temporal topological structures, a more general question concerns the topology of the full spatiotemporal Floquet band structure itself and its underlying symmetries. 
A material-transformation approach enables the construction of an adjoint system and the evaluation of a biorthogonal non-Abelian Zak phase, whose quantized change is accompanied by band inversion and an exchange of eigenstate symmetries \cite{gao2025topological}. 
For traveling-wave space-time photonic crystals, where spatial and temporal periodicities are intrinsically coupled, a complementary approach employs Lorentz transformations to a reference frame in which the modulation becomes purely spatial or temporal. 
In this frame, a conserved joint parity-time-reversal symmetry enforces the quantization of a spatiotemporal Zak phase in both subluminal and superluminal regimes \cite{caballero2026interface}. 
The resulting bulk topological invariant predicts interface states between crystals of distinct topological phases, while the orientation of the interface in space-time determines their characteristic dynamics, including propagation along moving boundaries, temporal localization, and frequency- or momentum-converted replicas.

Beyond their linear band structures and topological properties, space-time crystals also provide a means of controlling parametric instabilities and their nonlinear evolution. 
In a photonic-crystal lattice whose dielectric layers are modulated periodically in phase, the underlying spatial dispersion selects which positive- and negative-frequency modes are coupled by the temporal drive. 
The resulting momentum gaps provide parametric gain, while spatial phase control shifts these gaps asymmetrically and enables directional radiation at different oscillation frequencies \cite{lee2021parametric}. 
Kerr nonlinearity can stabilize the unstable modes of a mixed energy-momentum gap, producing an event soliton through the balance among energy-gap opening, momentum-gap opening, and nonlinearity. 
Such a spatiotemporal state is characterized by spatial confinement and temporal transience, thereby unifying previously distinct frameworks of Bragg and $k$-gap soliton formation \cite{zhang2026event}.

\subsection{Space-time metasurfaces}

\begin{figure}[!t]
    \centering
    \includegraphics[width=8.6cm]{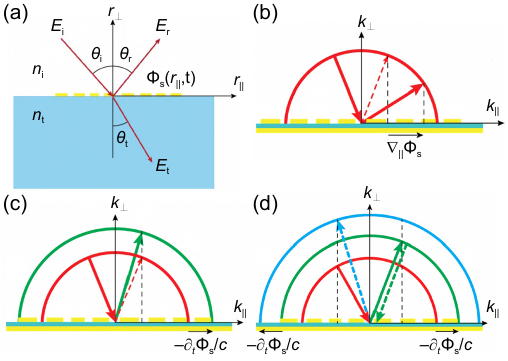}
    \caption{(a) Schematic of a light beam incident on a space-time gradient metasurface \captioncite{shaltout2015time}. 
    (b) Light reflected from a space-gradient metasurface acquires additional tangential momentum. 
    The dashed red line represents the reflected beam without the metasurface \captioncite{shaltout2015time}. 
    (c) Light reflected from a time-gradient metasurface shifts to a different isofrequency contour and acquires a frequency shift \captioncite{shaltout2015time}. 
    (d) Breaking of time-reversal in a time-gradient metasurface \captioncite{shaltout2015time}. 
    Panels (a)-(d) are reproduced from Ref.~\captioncite{shaltout2015time}, \textcopyright~2015 Optical Society of America (OSA Open Access Publishing Agreement).}\label{figure.19}
\end{figure}

Space-time metasurfaces and related dynamically reconfigurable metasurface platforms have recently attracted growing interest as approaches to high-speed wavefront control at optically thin interfaces \cite{zhang2018space, shaltout2019spatiotemporal,mikheeva2022space,zhong2024gigahertz}. 
A space-time metasurface may be viewed as an optically thin interface, as shown in Fig. \ref{figure.19}(a), which imposes a local phase discontinuity $\Phi_{\mathrm{s}}(r_{\| },t)$ on a reflected or transmitted wave, where $\mathrm{s=r,t}$, and the subscript $\| $ denotes the component parallel to the metasurface. 
Within the eikonal approximation, the spatial and temporal derivatives of the total phase determine the local wavevector and frequency. 
The incident and scattered waves therefore satisfy \cite{shaltout2015time}
\begin{equation}
    \begin{aligned}
    k_{\| , \mathrm{s}}&=k_{\| , \mathrm{i}}+\partial_{\|}\Phi_{\mathrm{s}},\\
    \omega_{\mathrm{s}}&=\omega_{\mathrm{i}}-\partial_{t} \Phi_{\mathrm{s}}.
    \end{aligned}
    \label{eq69}
\end{equation}
One sees that the spatial phase gradient shifts tangential momentum, whereas the temporal phase gradient exchanges energy with the wave. 
Together with the dispersion relation $k_{\mathrm{s}}=n_{\mathrm{s}}\omega_{\mathrm{s}}/c$, generalized laws of reflection and refraction can be given as
\begin{equation}
k_{\mathrm{s}} \sin \theta_{\mathrm{s}}-k_{\mathrm{i}}\sin\theta_{\mathrm{i}}= \partial_{\|} \Phi_{\mathrm{s}},
\label{eq70}
\end{equation}
evaluated at the frequency shifted by $-\partial_{t} \Phi_{\mathrm{s}}$. 
The corresponding momentum-space picture is illustrated by the change from a spatial to a temporal gradient. 
As illustrated in Figs. \ref{figure.19}(b)-(c), a purely spatial gradient displaces the tangential wavevector while keeping the scattered wave on the original isofrequency contour, while a temporal gradient transfers it to a different isofrequency contour and hence also changes the normal momentum component. 
A direct consequence of such a frequency shift is the breaking of time-reversal symmetry in Fig. \ref{figure.19}(d). 
The changes in the isofrequency contours of the incident and reverse beam are additive, and hence the wavenumber of the reverse beam is further increased compare to the incident one, leading to nonreciprocity. 

The generalized scattering laws above describe the frequency and momentum exchanged with an arbitrary space-time modulation. 
When the modulation is periodic, these exchanges are restricted to integer multiples of the modulation frequency and wavevector, giving rise to discrete Floquet channels.
Consider a metasurface whose surface capacitance is modulated by a traveling wave
\begin{equation}
C(z,t)=C_0+\Delta C \cos (Gz-\Omega t),
\label{eq71}
\end{equation}
where $G$ and $\Omega$ are the spatial and temporal modulation frequencies. 
An incident state $(k_{z,0},\omega_0)$ is then coupled to harmonics $k_{z,n}=k_{z,0}+nG$, $\omega_n=\omega_0+n\Omega$, and the normal wavevector is given by $k_{\bot ,n}=\sqrt{(\omega_n/c)^2-k_{z,n}^2}$. 
A mode is radiative when $k_{\bot ,n}$ is real and evanescent when it is imaginary.
Expanding the reflected and transmitted fields in Floquet harmonics and imposing the metasurface interface conditions yields an infinite set of coupled equations,
\begin{equation}
\sum_m M_{nm} \bm{a_m}=\bm{b_n},
\label{eq72}
\end{equation}
where $\bm{a_m}$ collects the reflected and transmitted amplitudes of the $m$-th Floquet order, $\bm{b_n}$ represents the incident field, and $M_{nm}$ describes the coupling between them.
For weak sinusoidal modulation, the problem can be truncated to $n=0,\pm 1$. 
Solving this coupled boundary-value problem determines the reflection and transmission amplitudes of the Floquet orders and reveals resonances associated with surface and leaky modes \cite{hadad2015space}. 

\begin{figure}[!b]
    \centering
    \includegraphics[width=17cm]{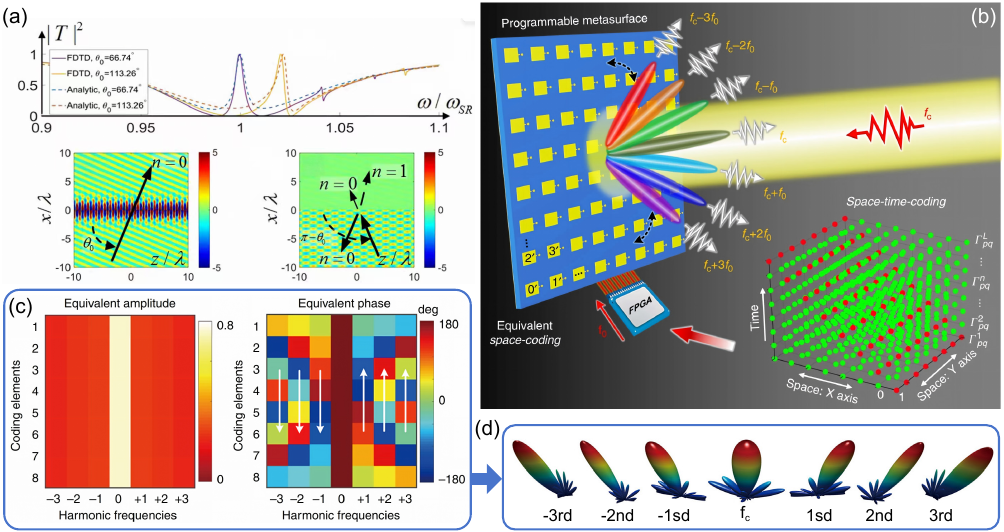}
    \caption{(a) Nonreciprocal response in a space-time gradient metasurface. 
    The metasurface is transparent for a wave incident from a certain direction, while being opaque for the counterpropagating one \captioncite{hadad2015space}. 
    (b) Illustration of a space-time-coding digital metasurface. The scattering coefficients can be dynamically controlled by the individually applied voltages to the PIN diodes \captioncite{zhang2018space}. 
    (c) Equivalent amplitude and phase distributions for the phase modulation scheme, which leads to harmonic beam steering in (d) \captioncite{zhang2018space}. 
    Panel (a) is reproduced with permission from Ref.~\captioncite{hadad2015space}, \textcopyright~2015 American Physical Society;
panels (b)-(d) are reproduced with permission from Ref.~\captioncite{zhang2018space}, \textcopyright~The Author(s) 2018 (CC BY 4.0)}\label{figure.20}
\end{figure}

This framework underlies a broad range of functionalities. 
A representative is the resonant nonreciprocal transmission, as depicted in Fig. \ref{figure.20}(a). 
In the space-time gradient metasurface considered by Hadad et al. \cite{hadad2015space}, spatial modulation enables coupling to a narrow leaky-mode resonance, producing an EIT-like transmission channel that remains reciprocal in the absence of temporal modulation. 
Adding a traveling temporal modulation lifts the degeneracy between counterpropagating incidence directions, so that at a chosen frequency one incident wave efficiently excites the leaky mode and is transmitted, while its counterpart is reflected. 
Related implementations include unequal emission and absorption patterns \cite{hadad2016breaking}, and optical nonreciprocal reflection produced by pump-induced space-time phase modulation \cite{guo2019nonreciprocal}. 
Beyond selecting the primary resonant channels, the modulation waveform itself can be engineered to distribute energy and phase among higher Floquet orders. 
In space-time-coding digital metasurfaces \cite{zhang2018space}, the cyclic binary temporal sequence applied to each element determines the Fourier amplitudes of the generated harmonics [see Fig. \ref{figure.20}(b)], and the spatial arrangement and relative time delays of these sequences establish harmonic-dependent phase profiles as shown in Fig. \ref{figure.20}(c). 
This enables simultaneous control of the scattered spectrum and wavefront, including harmonic beam steering in Fig. \ref{figure.20}(d). 
The implementation of multibit space-time gradients further enables dynamically reconfigurable nonreciprocal reflection \cite{zhang2019breaking}, high-efficiency frequency conversion via optimized modulation waveforms \cite{dai2020high}, and generation of  frequency-modulated continuous waves by a space-time-coding metasurface with nonlinearly periodic phases \cite{ke2022frequency}. 
Electrically driven optical metasurfaces further translate this approach to telecommunication wavelengths, where temporal voltage waveforms generate programmable sideband spectra and spatially delayed modulation independently steers selected frequency components \cite{sisler2024electrically}. 

Moreover, the Floquet formulation in Eq. (\ref{eq72}) naturally turns a space-time metasurface into a multiport system, in which each propagating channel serves as an external port, and evanescent harmonics provide internal degrees of freedom that help satisfy the local interface conditions. 
\begin{figure}[!t]
    \centering
    \includegraphics[width=17cm]{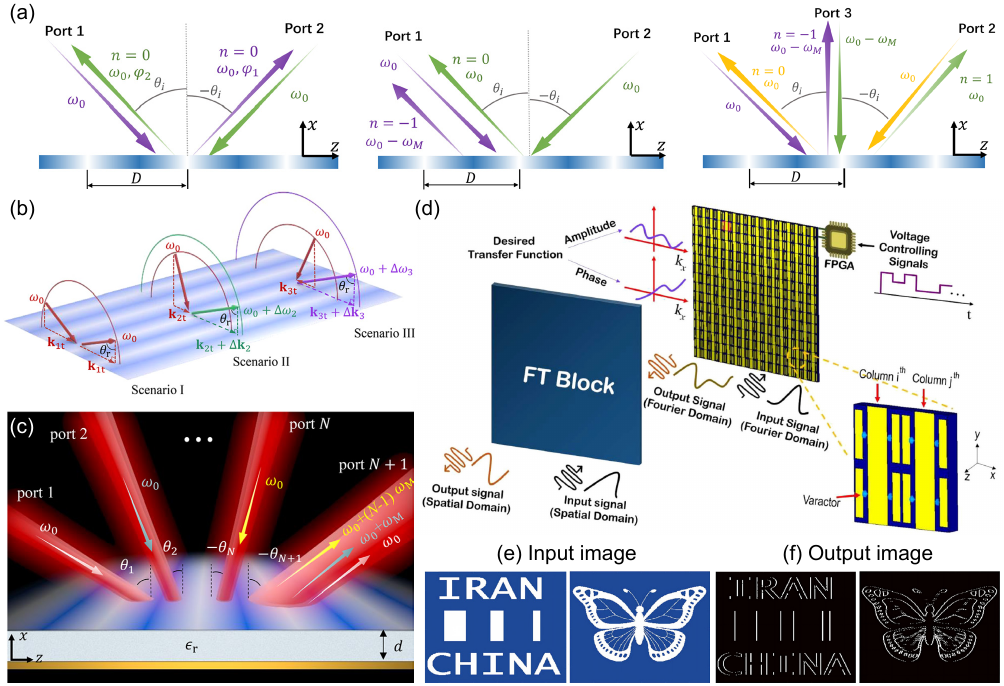}
    \caption{(a) By changing the applied modulation, a single metasurface can operate as an isolator, a nonreciprocal phase shifter, a quasi-isolator, or a circulator \captioncite{wang2020theory}. 
    (b) Principle of power combining using space-time metasurfaces, which allows distinct scattering response for different incidence angles at the same output port in (c) \captioncite{wang2021space}. 
    (d) Schematic demonstration of analog signal processing by a space-time metasurface. 
    Illustration of edge detection using a space-time metasurface, where the input and output images are shown in (e) and (f), respectively \captioncite{rajabalipanah2021analog}. 
    Panel (a) is reproduced with permission from Ref.~\captioncite{wang2020theory}, \textcopyright~2020 American Physical Society;
panels (b) and (c) are reproduced from Ref.~\captioncite{wang2021space}, \textcopyright~The Author(s) 2021 (CC BY 4.0);
panels (d)-(f) are reproduced from Ref.~\captioncite{rajabalipanah2021analog}, \textcopyright~The Author(s) 2021 (CC BY 4.0).}\label{figure.21}
\end{figure}
Beyond calculating the scattering response of a specific modulation, one can design a multifunctional metasurface by reconfiguring the applied modulation, which allows a single metasurface to operate as an isolator, a nonreciprocal phase shifter, a quasi-isolator, or a circulator [see Fig. \ref{figure.21}(a)] \cite{wang2020theory}. 
While reprogramming the modulation switches the functionality of the metasurface, a fixed modulation can itself support different transformations for different incident ports. 
For example, in Fig. \ref{figure.21}(b), a proper spatial modulation allows the reflected waves to gain different spatial momenta for different incident angles, while temporal modulation adds different frequency shifts to them \cite{wang2021space}. 
Consequently, waves incident from different directions can be coupled through different Floquet orders into the same output direction, as shown in Fig. \ref{figure.21}(c). 
Extending this multiport picture from a finite set of incident directions to a continuous spectrum of transverse momenta turns the metasurface into a programmable wave processor. 
When an input field is mapped into its spatial Fourier spectrum, the response of a selected harmonic can be engineered to reproduce a desired transfer function, as illustrated in Fig. \ref{figure.21}(d) \cite{rajabalipanah2021analog}. 
The same platform can therefore be reprogrammed for spatial differentiation, integration, equation solving, and edge detection, extending Floquet synthesis from discrete port routing to continuous-spectrum wave processing [see Figs. \ref{figure.21}(e)-(f)].

The Floquet-scattering framework assumes periodic modulation and primarily considers classical electromagnetic waves. Several recent developments extend this picture by relaxing these assumptions or introducing additional degrees of freedom. These include asynchronous and transient modulation beyond globally synchronized periodic driving, angular-momentum control through structured modulation, and the manipulation of nonclassical light. 
\begin{figure}[!b]
    \centering
    \includegraphics[width=17cm]{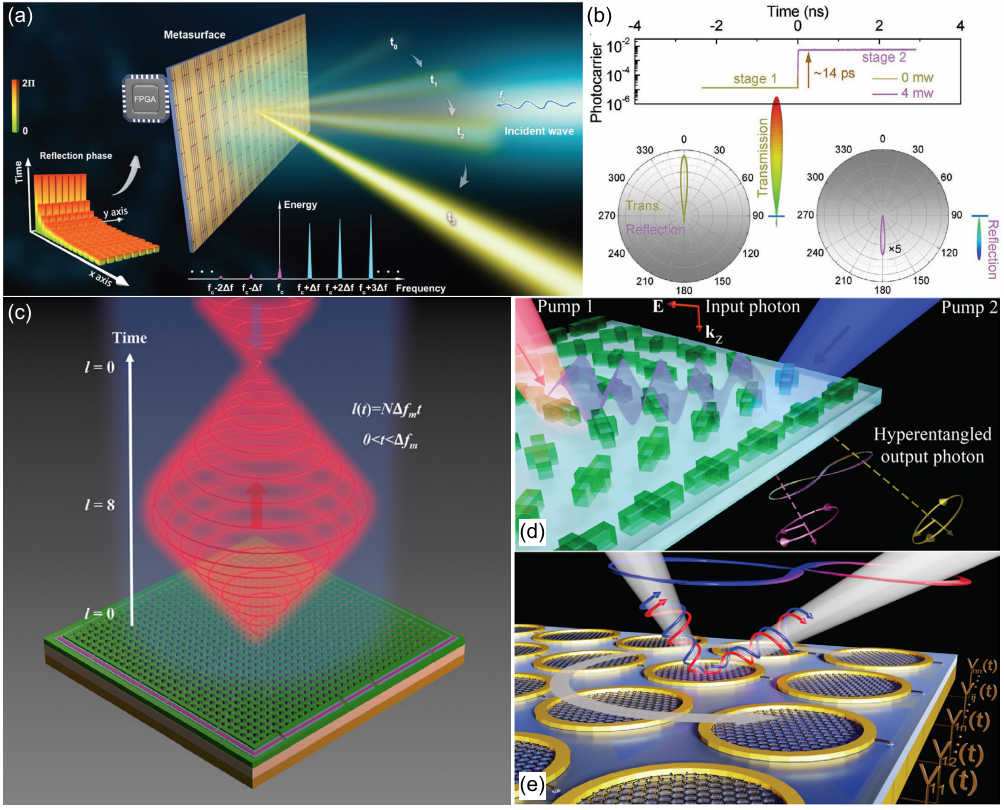}
    \caption{(a) Illustration of an asynchronous space-time-coding digital metasurface, enabling dynamic scattering control \captioncite{wang2022asynchronous}. 
    (b) Transient dynamics of photocarriers and the resulting ultrafast transition \captioncite{cong2020spatiotemporal}. 
    (c) Schematic of a time-varying optical vortex generated by a time-modulated metasurface with an azimuthal frequency gradient \captioncite{barati2020time}. 
    (d) Color-spin-path hyperentanglement and (e) generation of entangled vortex photon pairs by space-time metasurfaces \captioncite{kort2021space}. 
    Panel (a) is reproduced from Ref.~\captioncite{wang2022asynchronous}, \textcopyright~The Author(s) 2022 (CC BY 4.0);
panel (b) is reproduced with permission from Ref.~\captioncite{cong2020spatiotemporal}, \textcopyright~2020 Wiley-VCH;
panel (c) is reproduced from Ref.~\captioncite{barati2020time}, \textcopyright~The Author(s) 2020 (CC BY 4.0);
panels (d) and (e) are reproduced with permission from Ref.~\captioncite{kort2021space}, \textcopyright~2021 American Physical Society.}\label{figure.22}
\end{figure}
Beyond globally synchronized driving, asynchronous coding in Fig. \ref{figure.22}(a) assigns different modulation periods to individual meta-atoms, which introduces position-dependent frequency shifts, enabling time-varying interference scattering patterns with automatic temporal scanning \cite{wang2022asynchronous}. 
In contrast to a periodic modulation, optical pumping accesses a complementary transient regime, in which photocarrier generation and relaxation produce ultrafast switching and beam steering within a finite pump cycle rather than a Floquet steady state, resulting in reconfigurable beam steering in Fig. \ref{figure.22}(b) \cite{cong2020spatiotemporal, zhang2026space}. 
More recently, combining a patterned spatial grating with an optically induced transient grating has enabled reconfigurable spatiotemporal diffraction and spectral control \cite{stengel2026spatio}. 
Azimuthally structured modulation further extends space-time transitions from linear momentum to orbital angular momentum, correlating frequency sidebands with OAM states and enabling time-varying optical vortices and self-torqued beams in Fig. \ref{figure.22}(c) \cite{barati2020topological, barati2020time}. 
The synthetic-motion picture for space-time metamaterials can be extended to two-dimensional metasurfaces, where traveling-wave modulation produces nonreciprocal hyperbolic surface waves \cite{mazor2019one} and coherent multifrequency \v{C}erenkov radiation when it becomes superluminal \cite{oue2022vcerenkov}. 
Finally, space-time quantum metasurfaces bring these concepts into the quantum domain, allowing the spectral, spatial, and spin degrees of freedom of nonclassical light to be jointly controlled and enabling color-spin-path hyperentanglement and  generation of entangled vortex photon pairs, as illustrated in Figs. \ref{figure.22}(d)-(e) \cite{kort2021space}.

\section{Experimental platforms}

The experimental implementations of time-varying photonics require the relevant constitutive or effective parameters to be temporally modulated with sufficient speed, depth, coherence, and spatial uniformity over the interaction region. 
These requirements become particularly challenging at optical frequencies, where a nonadiabatic temporal change must occur on femtosecond timescales, while conventional nonlinearities are often either sufficiently fast but too weak, or sufficiently strong but limited by carrier, thermal, or structural relaxation \cite{saha2023photonic}. 
Resonant and nanostructured platforms can enhance the light-matter interaction and reduce the required modulation strength, but introduce trade-offs involving loss, bandwidth, damage threshold, and resonance lifetime \cite{galiffi2022photonics}. 
Nevertheless, recent experiments have shown promising potential for time-varying photonic systems via different strategies \cite{asgari2024theory}. 
Different experimental platforms address these requirements through distinct physical mechanisms. An important distinction is whether temporal modulation acts directly on the constitutive parameters of a propagating medium or on the effective Hamiltonian of an engineered wave system. The relevant modulation timescale, accessible coupling strength, dissipation, and degree of programmability then determine which temporal phenomena can be realized and observed. In the following, we review representative platforms from this perspective, emphasizing their operating principles, capabilities, and practical limitations.

\subsection{Electronically modulated RF and microwave platforms}

Electronically reconfigurable RF and microwave platforms provide a versatile means of engineering electromagnetic dispersion and controlling wave propagation through tunable circuit elements \cite{lim2005metamaterial,gil2006tunable,wang2024tunable}. 
Extending such tunability to explicit temporal modulation offers a practical route to realizing time-varying media, where distributed circuit parameters can be mapped onto effective constitutive parameters. 
As an example, we consider a TEM wave propagating along a transmission line \cite{caloz2005electromagnetic}. For an ideal lossless transmission line with time-dependent inductance $L(t)$ and capacitance $C(t)$ per unit length, the time-domain telegrapher's equations read
\begin{equation}
\begin{aligned}
\frac{\partial V}{\partial z}
&=-\frac{\partial[L(t)I]}{\partial t},\\
\frac{\partial I}{\partial z}
&=-\frac{\partial[C(t)V]}{\partial t},
\end{aligned}
\label{eq75}
\end{equation}
where $V$ and $I$ denote the voltage and current, respectively. These equations have the same mathematical structure as the one-dimensional Maxwell equations, with $V$ and $I$ corresponding to the electric and magnetic fields, and $C(t)$ and $L(t)$ playing the roles of permittivity and permeability, respectively. Consequently, a transmission line periodically loaded with subwavelength varactors can emulate a spatially homogeneous medium with an effective time-dependent permittivity $\varepsilon(t)$. For an ideal temporal interface, the continuity of charge $Q=CV$ and magnetic flux linkage $\Phi=LI$ corresponds to the continuity of $\bm{D}$ and $\bm{B}$ discussed in Section 2. In practice, these interface conditions may depend on the switching mechanism, while the loading elements must be modulated with nearly identical phase and amplitude, with the strong control signal isolated from the propagating probe \cite{moussa2023observation,xiong2025observation}.

\begin{figure}[!b]
    \centering
    \includegraphics[width=17cm]{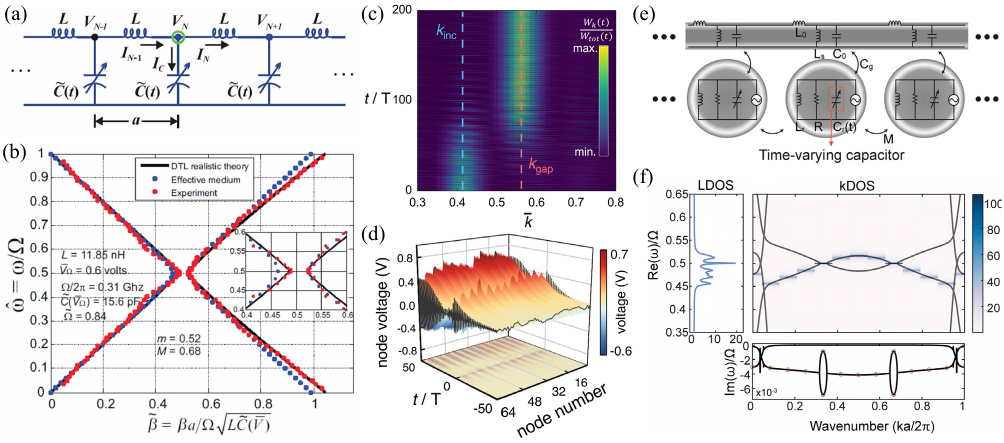}
    \caption{(a) A dynamic transmission line with time-varying capacitances \captioncite{reyes2015observation}. 
    (b) Measured band structures for the dynamic transmission line in (a) \captioncite{reyes2015observation}. 
    (c) Temporal evolution of the power spectrum in a dynamic transmission line under broadband excitation \captioncite{xiong2025observation}. 
    (d) Measured space-time field evolution in temporally cascaded photonic time crystals \captioncite{xiong2025observation}. 
    (e) Schematic of an effective photonic time crystal constructed by spatially concatenated time-varying LC resonators and a waveguide \captioncite{lee2026analogs}. 
    (f) Map of the momentum-resolved density of states, local density of states, and band structures \captioncite{lee2026analogs}. 
    Panels (a) and (b) are reproduced with permission from Ref.~\captioncite{reyes2015observation}, \textcopyright~2015 AIP Publishing LLC;
panels (c) and (d) are reproduced from Ref.~\captioncite{xiong2025observation}, \textcopyright~The Author(s) 2025 (CC BY-NC-ND 4.0);
panels (e) and (f) are reproduced with permission from Ref.~\captioncite{lee2026analogs}, \textcopyright~2026 American Physical Society.}\label{figure.23}
\end{figure}

Under continuous periodic driving, these circuits implement microwave Floquet media and photonic time crystals. 
The earliest dynamic-transmission-line experiments established the circuit-medium correspondence by observing a genuine momentum gap generated by periodic capacitance modulation, as illustrated in Figs. \ref{figure.23}(a)-(b) \cite{reyes2015observation}. 
A rcent experiment further resolves momentum-gap amplification and a temporal topological state by combining differential modulation, see Figs. \ref{figure.23}(c)-(d) \cite{xiong2025observation}. 
Varactor-loaded LC resonators embedded in metallic waveguides [see Fig. \ref{figure.23}(e)] enabled reconstruction of the complex non-Hermitian Floquet band structure, including exceptional transitions at the momentum-gap edges and noise-initiated oscillations under modulation \cite{park2022revealing}. 
Reducing the system to a modulated capacitive metasurface simplifies spatial synchronization and permits direct coupling to surface or free-space waves, demonstrating exponential surface-wave amplification \cite{wang2023metasurface}. 
Recent demonstrations in resonator and metasurface platforms further extend driven amplification to emission. 
Broadband circuit noise is used to map the momentum-gap LDOS and its analogue of spontaneous emission [see Fig. \ref{figure.23}(f)], followed by a threshold transition to parametric self-oscillation \cite{lee2026analogs}. 
A multiplier-driven modulation with an azimuthally varying modulation phase produces microwave vortex lasing carrying orbital angular momentum \cite{huang2026microwave}. 
Together, these experiments illustrate the progression from observing the Floquet band structure of a time-periodic medium to exploiting its dynamical instabilities for amplification and self-oscillation. The relatively accessible electronic modulation of microwave circuits makes them particularly suitable for systematically probing these effects.

\begin{figure}[!t]
    \centering
    \includegraphics[width=17cm]{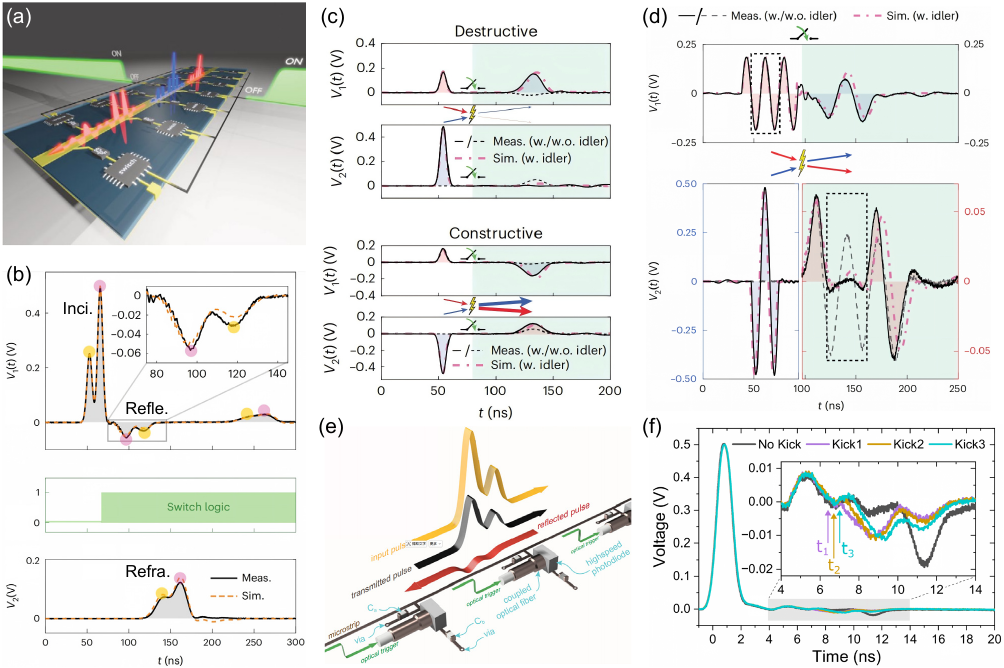}
    \caption{(a) Temporal interface based on a dynamic transmission line \captioncite{moussa2023observation}. 
    (b) Experimental observation of temporal reflection with an asymmetric input pulse \captioncite{moussa2023observation}.
    (c) Experimental demonstration of destructive and constructive photonic collisions \captioncite{galiffi2023broadband}. 
    (d) Ultrafast pulse shaping using a temporal interface \captioncite{galiffi2023broadband}. 
    (e) Temporal interface based on a microstrip transmission line triggered by optical pulses \captioncite{jones2024time}. 
    (f) Output voltage with the time interfaces being triggered at different times \captioncite{jones2024time}. 
    Panels (a) and (b) are reproduced with permission from Ref.~\captioncite{moussa2023observation}, \textcopyright~2023 Springer Nature;
panels (c) and (d) are reproduced with permission from Ref.~\captioncite{galiffi2023broadband}, \textcopyright~2023 Springer Nature;
panels (e) and (f) are reproduced from Ref.~\captioncite{jones2024time}, \textcopyright~The Author(s) 2024 (CC BY-NC-ND 4.0).
}\label{figure.24}
\end{figure}

To observe transient dynamics, the sinusoidal modulation is replaced by one or several rapid switching events. 
In a switched transmission-line metamaterial [see Fig. \ref{figure.24}(a)], additional capacitors are synchronously connected to or removed from the line through voltage-controlled switches, abruptly changing the distributed capacitance and wave impedance without requiring the capacitors themselves to be continuously tuned. 
Such switching produces large parameter changes within a few nanoseconds and enabled the observation of temporal reflection and broadband frequency translation at a temporal interface [see Fig. \ref{figure.24}(b)] \cite{moussa2023observation}. 
The same scheme also enables broadband coherent wave control by temporal interface. 
With two counterpropagating inputs, the relative phase and amplitude at the switching instant can control interference between time-reflected and time-refracted components, enabling broadband cancellation, enhancement, and pulse reshaping through photonic collisions \cite{galiffi2023broadband}, as shown in Figs. \ref{figure.24}(c)-(d). 
A faster modulation is achieved by employing an optical pulse to synchronously trigger high-speed photodiodes distributed along a microstrip in Fig. \ref{figure.24}(e), producing an almost instantaneous interface for microwave pulses, where waveform reversal, frequency translation, and phase conjugation are realized [see Fig. \ref{figure.24}(f)] \cite{jones2024time}. 

\subsection{Optical materials and metamaterials}

\begin{figure}[!b]
    \centering
    \includegraphics[width=17cm]{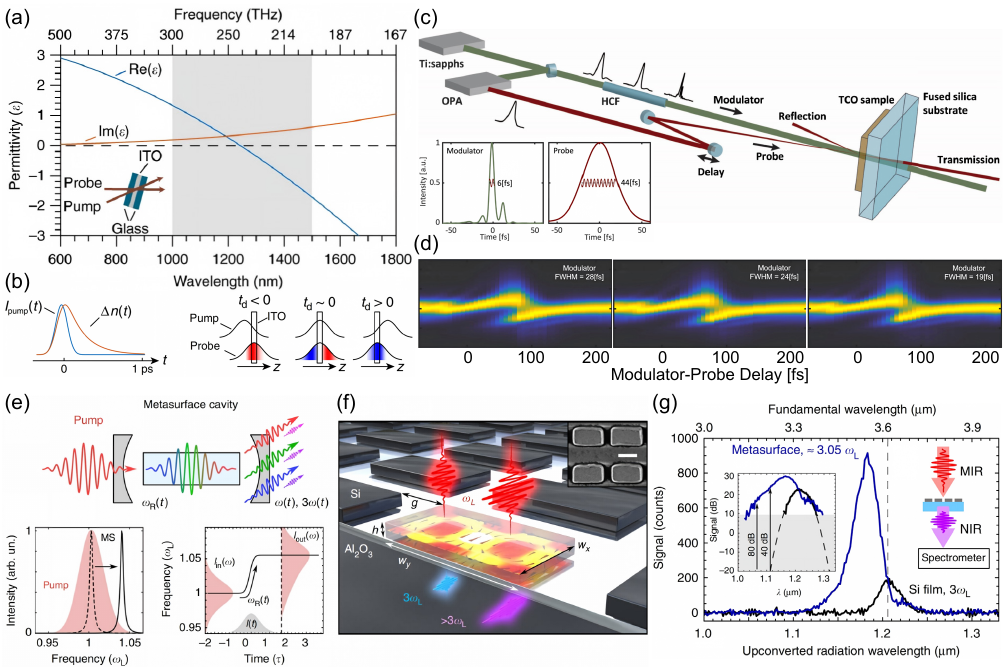}
    \caption{(a) The permittivity of an ITO film versus wavelength. Inserted figure shows a simplified experimental setup \captioncite{zhou2020broadband}. 
    (b) Illustration of the temporal index change of ITO excited by a pump pulse. The frequency shift of the probe is determined by the delayed time of pump pulse \captioncite{zhou2020broadband}. 
    (c) Experimental setup for time-refraction in the single-optical-cycle regime. Inserted figures show the intensity profile of the modulation and probe pulse, respectively. The red wiggly line represents the wave oscillations of the probe \captioncite{lustig2023time}. 
    (d) Transmission spectrum of the probe pulse in (c) for modulation pulses with different temporal widths \captioncite{lustig2023time}. 
   (e) Concept of blueshifted harmonic generation using the time-dependent metasurface shown in (f). 
   The pump changes the resonance spectrum of the metasurface, blueshifts the photons trapped inside, and subsequently upconverts the photons through conventional third-harmonic generation \captioncite{shcherbakov2019photon}. 
   (g) Comparison of the third harmonic generation in the metasurface and an unstructured Si film \captioncite{shcherbakov2019photon}. 
   Panels (a) and (b) are reproduced from Ref.~\captioncite{zhou2020broadband}, \textcopyright~The Author(s) 2020 (CC BY 4.0);
panels (c) and (d) are reproduced from Ref.~\captioncite{lustig2023time}, \textcopyright~The Author(s) 2023 (CC BY 4.0);
panels (e)-(g) are reproduced from Ref.~\captioncite{shcherbakov2019photon}, \textcopyright~The Author(s) 2019 (CC BY 4.0).}\label{figure.25}
\end{figure}

At optical frequencies, the most direct way to realize time-varying photonics is to use an ultrafast pump to modify the electronic response of an active material. 
Interband and intraband photocarrier excitation can produce comparatively large sub-picosecond changes, whereas virtual-state nonlinearities are generally faster but weaker. 
Transparent conducting oxides such as indium tin oxide (ITO) and aluminum-doped zinc oxide (AZO) are particularly suitable because pump-induced carrier redistribution shifts the plasma frequency of their Drude response, while the small refractive index near the epsilon-near-zero point converts this shift into a large relative index modulation \cite{alam2016large, caspani2016enhanced, guo2016large}, as illustrated in Figs. \ref{figure.25}(a)-(b). 
These pump-induced changes also illustrate the close connection between nonlinear optics and time-varying photonic media. Although the modulation may originate from nonlinear light-matter interactions, the response of a sufficiently weak probe can be described by a linear wave equation with material parameters determined by the pump. In this description, the pump-induced temporal variation governs the probe dynamics, including frequency conversion and time refraction. However, the distinction is not absolute. Optically induced time refraction may also be described in terms of cross-phase modulation, while conventional nonlinear frequency-mixing processes can coexist with time-varying propagation \cite{zhou2020broadband,asgari2024theory}. 
More recently, parametric amplification dominated by time refraction is demonstrated under strong optical pumping in ENZ materials, highlighting the role of dynamically varying material properties in nonlinear optical processes \cite{vaddi2026parametric}. 
In realistic optical materials, the pump-induced dielectric response is generally dispersive, with its temporal evolution governed by electronic excitation, relaxation, and recovery. The corresponding description therefore involves a dispersive constitutive response $\varepsilon(\omega,t)$ rather than only an instantaneous nonlinear susceptibility. 
When this variation is sufficiently slow, the probe primarily accumulates a time-dependent phase, leading to photon acceleration, spectral translation, and time refraction \cite{zhou2020broadband, pang2021adiabatic, liu2021photon}.

The same pump-probe configuration enters a qualitatively different regime when the modulation approaches a single optical cycle. 
The accumulated-phase picture then becomes insufficient, and the response must be treated as a nonadiabatic scattering process. 
Increasing the modulation speed progressively broadens the generated spectrum and enhances coupling between positive- and negative-frequency channels, giving rise to phase conjugation, time reversal, and ultimately time reflection as shown in Figs. \ref{figure.25}(c)-(d) \cite{vezzoli2018optical, bohn2021spatiotemporal, lustig2023time}. 
Multiple ultrafast modulation windows further form temporal interferometers, as illustrated by double-slit time diffraction \cite{tirole2023double}. 
In addition, time-dependent material responses can also modify conventional nonlinear frequency-conversion processes. For example, in second-harmonic generation at an optically pumped ITO interface, the observed modulation arises not only from changes in the fundamental field but also from a time-dependent second-order nonlinear susceptibility \cite{tirole2024second}. 
Recent demonstrations of optical time reflection at a single temporal interface provide essential building blocks for optical PTCs \cite{segal2026time}. 
These optical demonstrations also raise a fundamental question concerning the microscopic origin of such ultrafast material responses. In transparent conducting oxides, optical pumping redistributes conduction electrons, modifying their effective mass and hence the plasma frequency through the nonparabolic electronic band structure. Recent experimental and theoretical studies have investigated how this process evolves and saturates under intense optical excitation \cite{li2026ultrafast}. Meanwhile, microscopic density-matrix approaches have provided further insight into the interplay between electronic thermalization, quantum coherence, and the optical response under single-cycle excitation \cite{un2026origin}. Nevertheless, certain features of the ultrafast relaxation observed in optical time-refraction experiments remain incompletely understood \cite{lustig2023time}. Establishing a connection between microscopic nonequilibrium dynamics and the effective time-dependent constitutive response is therefore an important direction toward realistic optical temporal interfaces and photonic time crystals.

On the other hand, resonantly enhanced metastructures provide a complementary strategy for strengthening these effects without relying exclusively on a larger intrinsic material modulation [see Figs. \ref{figure.25}(e)-(g)]. 
Plasmonic-epsilon-near-zero (plasmonic-ENZ) hybrid structures enhance the interaction by improving coupling to an ultrathin active layer and concentrating the field within it \cite{alam2018large, bruno2020negative, pang2021adiabatic, bohn2021all1}, whereas dielectric metasurfaces exploit Mie-type or quasi-BIC resonances with lower absorption and generally greater tolerance to the intense optical pumping \cite{shcherbakov2019photon, karl2020frequency}. 
In such structures, the response is formulated in terms of a time-dependent resonance frequency, linewidth, and external coupling rate, complementing the bulk constitutive description with time-dependent coupled-mode theory. 
This framework captures resonantly enhanced frequency conversion, spectral interference, harmonic shifting, and space-time scattering \cite{shcherbakov2019photon, guo2019nonreciprocal, karl2020frequency}. 
The enhancement is nevertheless accompanied by a trade-off, i.e., a higher quality factor strengthens the light-matter interaction but narrows the usable bandwidth and increases the resonance lifetime, which may become incompatible with an ultrashort modulation pulse. 
Furthermore, multilayered ENZ structures, polaritonic metasurfaces, and two-dimensional materials can be viewed as extensions of the same strategy, combining active material dynamics with engineered modal confinement \cite{ge2015dynamical, verre2019transition, indukuri2019ultrasmall, indukuri2020ws2, maccaferri2021enhanced, mann2021ultrafast}. 
In addition, it should be noted that the large refractive-index modulation near material resonances is often accompanied by enhanced absorption, so the experimentally accessible frequency-conversion efficiency or net amplification must be evaluated by considering both the modulation strength and the associated losses.

\subsection{Synthetic dimensions}

Synthetic platforms replace the ultrafast modulation of a material parameter by the programmable modulation of an effective lattice Hamiltonian. 
Internal states of atoms or photons serve as lattice sites, while electronic or optical control signals vary the coupling amplitude, coupling phase, onsite potential, or gain and loss of the Hamiltonian. 
The relevant switching speed is therefore set by modulators and driving fields rather than by the response time of an optical material, providing a flexible route to construct temporal interfaces and periodically driven Hamiltonians at optical carrier frequencies. 
We emphasize that a temporal interface in these implementations corresponds to a change in the effective evolution Hamiltonian, rather than necessarily requiring an abrupt change in the electromagnetic constitutive parameters. The resulting dynamics can reproduce the modal scattering and Floquet phenomena discussed in the preceding sections, while the corresponding state variables and interface conditions are determined by the effective model.

\begin{figure}[!t]
    \centering
    \includegraphics[width=15cm]{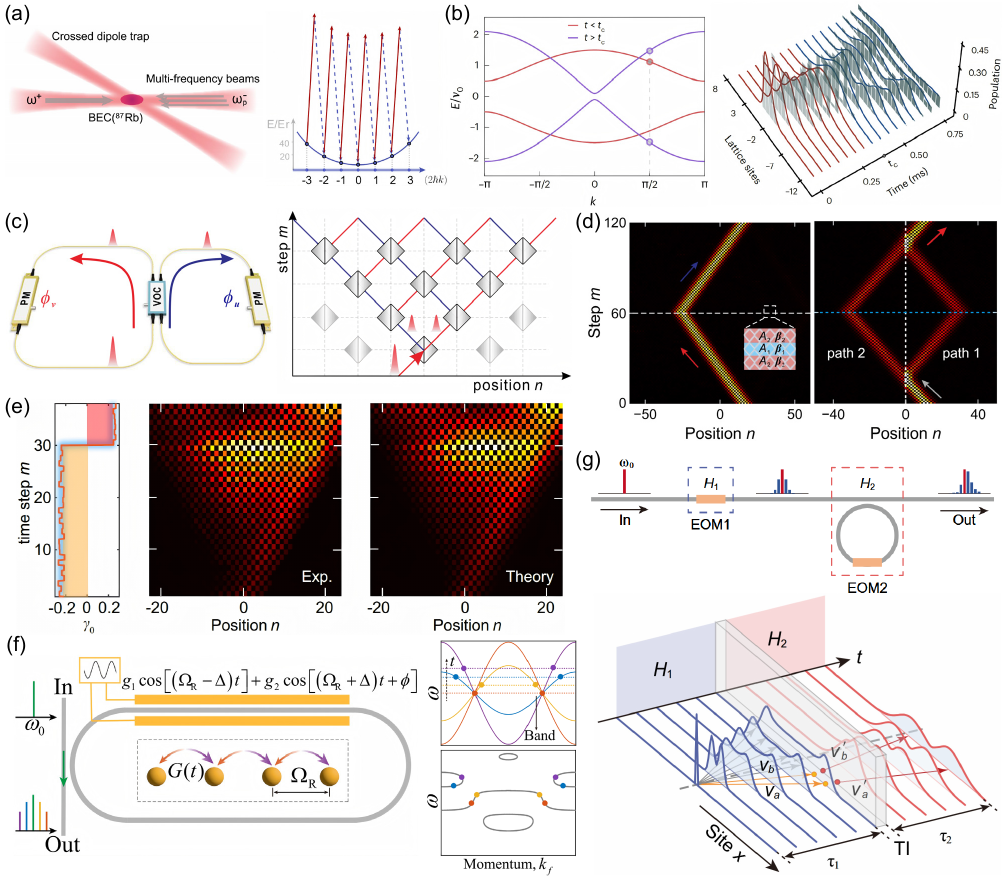}
    \caption{(a) Synthetic momentum lattice based on ultracold atoms \captioncite{dong2024quantum}. (b) Band structures before and after the temporal interface and evolution of the wave function \captioncite{dong2024quantum}. (c) Schematic of a time-varying optical system \captioncite{ye2023reconfigurable}. (d) Time reversal using a temporal superlens and a temporal Mach-Zehnder interferometer in a synthetic temporal lattice \captioncite{ye2023reconfigurable}. (e) Temporal topological edge state in a synthetic temporal lattice \captioncite{ren2025observation}. (f) Schematic of a time-periodic system based on synthetic frequency dimension, and evolution of band trajectories, leading to a dynamic band structure \captioncite{ye2025construction}. (g) Schematic of a temporal interface based on synthetic frequency dimension with the evolution of the wave function \captioncite{he2025observing}. 
    Panels (a) and (b) are reproduced with permission from Ref.~\captioncite{dong2024quantum}, \textcopyright~2023 Springer Nature;
panels (c) and (d) are reproduced from Ref.~\captioncite{ye2023reconfigurable}, \textcopyright~The Author(s) 2023 (CC BY-NC-ND 4.0);
panel (e) is reproduced from Ref.~\captioncite{ren2025observation}, \textcopyright~The Author(s) 2025 (CC BY-NC-ND 4.0);
panel (f) is reproduced with permission from Ref.~\captioncite{ye2025construction}, \textcopyright~2025 American Physical Society;
panel (g) is reproduced from Ref.~\captioncite{he2025observing}, \textcopyright~The Author(s) 2025 (CC BY-NC 4.0).}\label{figure.26}
\end{figure}

For example, a synthetic lattice can be constructed by the discrete momentum states of a Bose-Einstein condensate in ultracold atoms [see Fig. \ref{figure.26}(a)] \cite{dong2024quantum}. 
Adjacent states are coupled through two-photon Bragg transitions driven by multifrequency laser beams. 
The RF drive applied to the acousto-optic modulator generates a multifrequency laser  beams incident on the Bose-Einstein condensate, which controls the corresponding hopping amplitude. 
This platform allows individual hopping terms, disorder, and potentially many-body interactions to be controlled in a flexible manner. 
A temporal interface can then be implemented by rapidly changing the coupling amplitude, producing simultaneous quantum time reflection and refraction [see Fig. \ref{figure.26}(b)], which are robust against considerable disorder. 
Varying the switching duration further connects the sudden and adiabatic regimes. 

Moreover, the synthetic temporal lattice in coupled-loop fiber-based platforms provides another convenient way to study the dynamics in time-varying systems, where optical pulses circulate through two fiber loops with different lengths [see Fig. \ref{figure.26}(c)]. 
The relative arrival time of each pulse labels the synthetic lattice position $n$, and the number of round trips acts as the discrete evolution step $m$. 
The evolution of the pulses is described by \cite{regensburger2011photon, regensburger2012parity,  ye2023reconfigurable, hu2024observing}
\begin{equation}
  \begin{aligned}
  u^{m+1}_{n}=\left[\cos (\beta) u^{m}_{n+1}+i\sin (\beta)v^{m}_{n+1}\right] e^{i\phi_u},\\
  v^{m+1}_{n}=\left[\cos (\beta) v^{m}_{n-1}+i\sin (\beta)u^{m}_{n-1}\right] e^{i\phi_v},
  \end{aligned}\label{eq73}
\end{equation}
where $u^{m}_n$ ($v^{m}_n$) denote the pulse amplitude in the short (long) loop at the $m$-th step with a position index $n$. 
$\beta$ is the coupling angle of the tunable optical coupler, which controls the hopping strength between the two sublattices.
The phase modulators impose programmable hopping phases $\phi_u$ and $\phi_v$. 
These elements can be driven by arbitrary waveform generators, so their values may be switched at a selected evolution step or programmed as functions of both $n$ and $m$. 
In particular, phase shifts $\phi_u$ and $\phi_v$ introduce the effective scalar potential $\phi$ and vector potential $A$ \cite{ye2023reconfigurable}
\begin{equation}
\phi=\frac{\phi_v+\phi_u}{2}, \hspace{0.5cm} A=\frac{\phi_v-\phi_u}{2}.
\label{eq74}
\end{equation}
The main feature of this platform is its step-by-step programmability, where coupling, phase, gain, and loss for each site can be controlled independently at each time step. 
This enables reconfigurable scalar- and vector-potential interfaces, which lead to temporal total internal reflection, Klein tunnelling, temporal superlens, and temporal Mach-Zehnder interferometers [see Fig. \ref{figure.26}(d)] \cite{ye2023reconfigurable}. 
Space-time-dependent modulation waveforms further generate moving gauge-potential barriers, resulting in velocity-dependent discrete refraction \cite{qin2024observation}. 
By programming the coupling and gain-loss sequence periodically, the same architecture can realize temporal topological boundary states [see Fig. \ref{figure.26}(e)] \cite{ren2025observation}, temporal scattering processes from switching gain-loss-induced Dirac masses \cite{yu2024dirac}, and localized states at a space-time event \cite{feis2025space}.

In addition, the synthetic frequency lattices discussed in Section 3.5 can be mapped beyond a static Hamiltonian to emulate the underlying physics in time-varying systems. 
In an optical ring resonator, the resonant modes act as lattice sites, and electro-optic modulation at frequency $N \Omega_{\mathrm{R}}$ couples frequency modes separated by $N$ sites, where $\Omega_{\mathrm{R}}$ is the free spectral range of the ring. 
The modulation frequency therefore selects the hopping range, its amplitude controls the coupling strength, and its phase supplies the hopping phase or synthetic gauge potential \cite{yuan2021synthetic}, which provides a direct correspondence between the electrical drive waveform and the synthetic Hamiltonian. 
As a simple example, in the integrated thin-film lithium-niobate resonator in Fig. \ref{figure.26}(f), two modulation tones applied to the electrodes at frequencies oppositely detuned from the free spectral range create a time-dependent coupling through their beating. 
The amplitudes, phases, and detunings of the RF tones can therefore program different time-varying Hamiltonians with different band trajectories, leading to the observation of distinct dynamic band structures. 
A different strategy is illustrated in Fig. \ref{figure.26}(g), where the Hamiltonians before and after the temporal interface are encoded by two spatially separated electro-optic modulators rather than by rapidly switching a single device. 
The input field first undergoes a finite evolution time interval in the synthetic space under the Hamiltonian $H_1$ when passing through the first modulator. 
Once it leaves the modulator, the coupling between frequency modes vanishes and the evolution in the synthetic frequency dimension is frozen until the field enters the ring. 
The evolution resumes when the field encounters the second modulator inside the ring, which implements the evolution under the Hamiltonian $H_2$. 
Consequently, although the two Hamiltonians are realized at different physical positions, no effective evolution time elapses between them. 
The transition from $H_1$ to $H_2$ represents an ideally abrupt temporal interface with an effectively infinitesimal switching timescale, without requiring ultrafast electronic switching. 
The modulation order, amplitude, and phase of the two modulators determine the hopping range, strength, and phase on either side of the interface, enabling band shifting, folding and unfolding, and effective two-dimensional temporal operations. 
A spectral tomography is employed to reconstruct the associated state projections, which verifies momentum conservation across the entire bands \cite{he2025observing}.

\subsection{Other wave systems}

Related phenomena in time-varying systems can also be simulated in other wave systems, including water, acoustic, and mechanical waves \cite{bacot2016time, kim2024temporal, li2025topological}. 
These systems can be generally described by a second-order equation
\begin{equation}
\frac{\partial}{\partial t}\left[\mathcal{M}(x,t)\frac{\partial\psi}{\partial t}\right]  + \mathcal{G}(x,t)\frac{\partial\psi}{\partial t} -\frac{\partial}{\partial x}\left[\mathcal{K}(x,t)\frac{\partial\psi}{\partial x}\right] + \mathcal{V}(x,t)\psi=S(x,t).
\end{equation}
Here, $\psi$ denotes the relevant wave field, $\mathcal{M}$ is an inertia or energy-storage parameter, $\mathcal{K}$ governs the spatial restoring force and wave coupling, $\mathcal{V}$ represents a local restoring term or resonance, and $\mathcal{G}$ describes dissipation or effective gain. 
In a homogeneous, lossless, and resonance-free medium, the propagation velocity follows $v^2
=\mathcal{K}/\mathcal{M}$, and the combination of $\mathcal{M}$ and $\mathcal{K}$ also determines the wave impedance. 
Under appropriate assumptions, this equation can be mapped onto the scalar form of Maxwell's equations through the correspondence $\mathcal{M}\rightarrow \varepsilon$, $\mathcal{K}\rightarrow \mu^{-1}$, and $\mathcal{G}\rightarrow \sigma$. 
Temporal modulation of inertia, stiffness, coupling, resonance, or gain and loss can therefore reproduce temporal-interface and Floquet dynamics similar to those in electromagnetic waves.

\begin{figure}[!t]
    \centering
    \includegraphics[width=17cm]{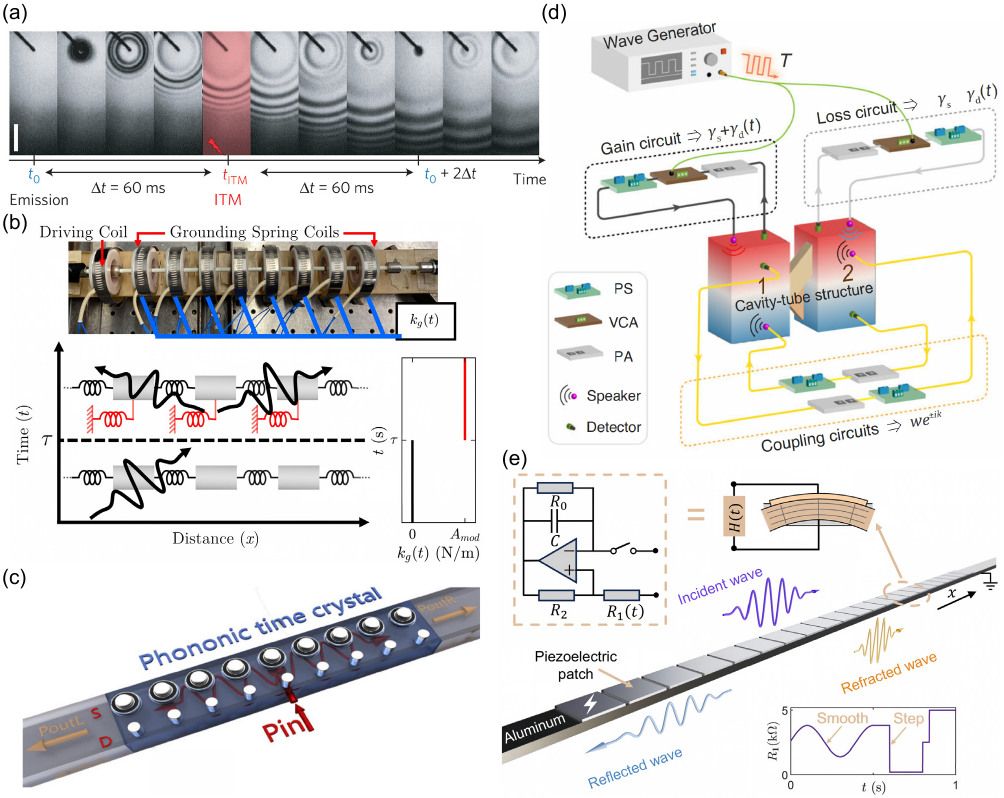}
    \caption{(a) Time reversal by using an instantaneous time mirror in water waves \captioncite{bacot2016time}. 
    (b) Temporal interface generated by an acoustic phononic lattice composed of acoustic meta-atoms \captioncite{kim2024temporal}. 
    (c) Schematic of phononic time crystals \captioncite{liu2026temporal}. 
    (d) Illustration of a pair of coupled acoustic cavities \captioncite{tong2025observation}. 
    (e) Schematic of a time-varying metabeam \captioncite{wang2025experimental}.
    Panel (a) is reproduced with permission from Ref.~\captioncite{bacot2016time}, \textcopyright~2016 Macmillan Publishers Limited;
panel (b) is reproduced with permission from Ref.~\captioncite{kim2024temporal}, \textcopyright~2024 American Physical Society;
panel (c) is reproduced from Ref.~\captioncite{liu2026temporal}, \textcopyright~The Author(s) 2026 (CC BY 4.0);
panel (d) is reproduced from Ref.~\captioncite{tong2025observation}, \textcopyright~The Author(s) 2025 (CC BY 4.0);
panel (e) is reproduced from Ref.~\captioncite{wang2025experimental}, \textcopyright~The Author(s) 2025 (CC BY 4.0).}\label{figure.27}
\end{figure}

Water waves provide a direct macroscopic realization. 
As illustrated in Fig. \ref{figure.27}(a), a vertical acceleration of the entire liquid bath abruptly changes the effective gravity and hence the gravity-capillary dispersion, which generates an instantaneous time mirror \cite{bacot2016time}. 
This produces a backward, time-reversed component that compensates dispersion and refocuses at the original source, enabling broadband time reversal and temporal holography. 
Note that similar time-reversal operations have previously been discussed for controlling acoustic and electromagnetic wave dynamics \cite{yariv1979compensation, miller1980time, fink1992time, fink2001acoustic, de2002overcoming, yanik2004time, pendry2008time, chumak2010all, sivan2011time, yuan2016time}. 
A more reconfigurable implementation uses electrostriction between a conductive water surface and an electrode placed above it \cite{apffel2022experimental}. 
The applied voltage modifies the wave velocity, and therefore short electrical pulses generate spatially uniform temporal interfaces. 
This scheme offers large modulation depth, short response time, and arbitrary pulse timing, allowing the construction of disordered time-varying media.

Acoustic and phononic platforms can also offer localized and programmable control of the effective Hamiltonian. 
In a discrete phononic lattice formed by repelling ring magnets in Fig. \ref{figure.27}(b), magnetic interactions provide the intersite coupling, while electromagnetic coils generate an electrically tunable grounding stiffness. 
A synchronous step in the coil current changes the lattice dispersion and group velocity, leading to temporal refraction and reflection, frequency conversion at conserved wavelength, and dynamically tunable slow sound \cite{kim2024temporal}. 
A complementary approach uses digitally virtualized acoustic meta-atoms in Fig. \ref{figure.27}(c) to synthesize a dispersive constitutive response \cite{liu2026temporal}. 
Each meta-atom consists of a microphone and speaker connected through a microcontroller, which performs a programmable convolution and generates a Lorentzian or anti-Lorentzian monopolar response. 
Periodically switching the resonant strength changes the effective compressibility, thereby realizing a phononic time crystal in an acoustic waveguide, where amplified transmission and strong emission enhancement within the momentum gap are experimentally demonstrated. 
Extending the temporal unit cell into an ABAC-type supercell folds the Floquet bands and opens additional momentum gaps, yielding dual-band transmission amplification. 
On the other hand, for coupled-acoustic-cavity systems in Fig. \ref{figure.27}(d), microphones, loudspeakers, and feedback circuits create time-dependent onsite gain and loss, while phase shifters in the coupling circuits synthesize the Bloch momentum. 
Periodic switching of balanced gain and loss realizes parity-time-symmetric Floquet lattices \cite{tong2025observation}. 
The Floquet operator can be reconstructed from the measured time-domain responses, which allows the resolution of both complex quasienergies and eigenstates, enabling direct observation of momentum-band inversion, topological invariants, and temporally localized interface states. 
By further introducing a position-dependent coupling modulation, a monoatomic acoustic space-time crystal can be realized, where oblique energy-momentum Brillouin zones and nontrivial $\pi$-gap edge states are observed \cite{tong2026acoustic}. 

Elastic and active mechanical systems extend these ideas to continuum engineering structures and explicitly non-Hermitian dynamics. 
In a time-varying metabeam shown in Fig. \ref{figure.27}(e), piezoelectric patches connected to programmable shunt circuits simultaneously sense the local bending field and act as mechanical actuators, which elongates and contracts in response to the applied voltage, thereby controlling its effective bending stiffness \cite{wang2025experimental}. 
Analog switches are employed to produce a spatially uniform stiffness step on a sub-microsecond timescale, which enables temporal refraction and reflection of flexural waves. 
Digital potentiometers generate smooth, periodic, or aperiodic profiles, demonstrating smart waveform morphing and information coding. 
A different mechanical implementation modulates non-Hermiticity rather than stiffness. 
In a chain of active rotational oscillators, velocity-dependent feedback torques serve as onsite gain or loss.
A spatially alternating gain-loss profile opens a Bloch momentum gap in a static configuration, while a single sign flip of the entire profile performs a temporal interface. 
The resulting transition converts a growing mode into a decaying mode and produces a topological temporal interface state localized around the switching event, which is robust against temporal imperfections \cite{li2025topological}.

These examples illustrate that the essential physics of time-varying wave systems is not restricted to electromagnetic media. Temporal modulation of inertia, stiffness, coupling, or gain and loss can generate analogous wave-scattering and Floquet phenomena across different physical platforms. Meanwhile, the relevant state variables, interface conditions, and material responses remain system-dependent. Such platforms therefore provide complementary opportunities to investigate the general principles of time-varying wave dynamics, while highlighting the importance of distinguishing universal mathematical structures from their specific physical realizations.

\section{Conclusions and outlook}
To summarize, we review how temporal modulation reshapes wave scattering, band structures, and light-matter interactions in photonic systems. 
We start from illustrating the basic physics of a single temporal interface, where the abrupt temporal change redistributes the incident wave among the modes supported by the new medium. 
This simple modal projection picture provides the foundation for understanding temporal interfaces, including temporal scattering, frequency conversion and phase conjugation. 
We then discuss the temporal scattering matrix as a general description of temporal scattering and extend the discussion to temporal slabs and multilayer structures, where waves scattered at successive interfaces interfere to enable various forms of coherent wave control.
Periodic repetition of the modulation also leads to the Floquet regime, where Floquet theory and transfer-matrix methods can be applied to reveal exotic wave dynamics in time-periodic systems. 
On the other hand, we show that synthetic frequency dimension provide a complementary lattice interpretation for photonic systems under dynamic modulation, in which modulation couples discrete frequency channels and enables programmable Hamiltonians, effective gauge fields, and higher-dimensional physics.
Finally, we discuss how joint modulation of a medium in space and time allows both frequency and momentum to be exchanged, leading to space-time interface effects, nonreciprocity, synthetic motion, luminal localization and amplification, and multiport wave processing in space-time metasurfaces. 
We also briefly summarize experimental demonstrations of these theoretical concepts across various platforms, highlighting recent developments in time-varying media ranging from single temporal interfaces to space-time wave engineering.

Despite rapid progress in this field of time-varying photonics, several challenges and open questions remain to be addressed.
The fundamental principles of wave propagation in time-varying media are relatively well established within the framework of linear electrodynamics. Temporal interface conditions, generalized conservation laws, and Floquet theory provide a coherent description of wave scattering, frequency conversion, and amplification. These concepts have been validated experimentally across diverse platforms.

From an experimental perspective, a central challenge is to achieve sufficiently strong and ultrafast modulation while maintaining low loss, spatial uniformity, and precise control over the modulation waveform. These requirements are particularly demanding at optical frequencies, where modulation is ultimately constrained by the underlying material response. In addition to modulation speed and depth, the efficiency of energy transfer from the external drive, dynamical stability, and noise become increasingly important, especially for systems involving parametric amplification. Addressing these challenges is essential for moving beyond proof-of-principle demonstrations toward precise and reliable control of electromagnetic wave dynamics in time-varying media. Such capabilities would provide the foundation for programmable time-varying photonic systems and enable more systematic comparisons between theoretical predictions and experimental observations.

At a more fundamental level, a major theoretical challenge is to establish a microscopic understanding of the ultrafast electromagnetic response of materials driven far from equilibrium. Most existing descriptions treat the time dependence of the constitutive parameters as externally prescribed, while in real materials it arises from nonequilibrium electronic and other microscopic dynamics. The limitations of our current understanding are already evident in experiments. For example, single-cycle time-refraction measurements in transparent conducting oxides have revealed unexpectedly fast refractive-index relaxation, whose microscopic origin remains unresolved \cite{lustig2023time}. Developing a predictive connection between microscopic material dynamics and macroscopic time-varying electrodynamics is therefore essential. Such a framework should consistently account for material dispersion, nonequilibrium relaxation dynamics, dissipation, and fluctuations, particularly when the modulation timescale approaches the intrinsic response times of the medium \cite{horsley2025macroscopic,sun2025formulation}.

Beyond material response, another fundamental question concerns the physical principles that constrain electromagnetic evolution in general time-dependent systems. While symmetries and conservation laws provide powerful tools for understanding idealized models, their implications for finite, dispersive, dissipative, and imperfectly modulated systems remain to be fully understood. In particular, the relationship between the bulk properties of time-periodic media and observable scattering dynamics, including the robustness of topological phenomena in realistic configurations, requires further investigation. Establishing which dynamical properties are determined by general electromagnetic principles, rather than by specific modulation protocols, could provide a more unified understanding of temporal interfaces, photonic time crystals, and space-time media.

Another frontier lies beyond linear wave dynamics, where temporal modulation interacts with nonlinearities and quantum fluctuations to produce new nonequilibrium phenomena. The interplay between temporal modulation and optical nonlinearities offers opportunities for exploring and controlling complex dynamical states, ranging from nonlinear gap solitons and spontaneous pattern formation in photonic time crystals to spatiotemporally localized event solitons in space-time crystals \cite{pan2023superluminal, kiselev2025symmetry, zhang2026event}. 
In addition, another approach based on externally driven optical resonators and lasers further illustrates the potential of temporal control for manipulating nonlinear dissipative states \cite{zou2025resonantly}. At the quantum level, temporal scattering and parametric amplification offer opportunities for photon-pair generation, squeezing, and quantum-state engineering, while raising fundamental questions about the roles of material fluctuations and noise \cite{mirmoosa2025quantum,sustaeta2025quantum}. Understanding how nonlinear interactions and quantum fluctuations modify the instabilities and scattering processes of time-varying systems may uncover new physical mechanisms without direct counterparts in conventional static photonic systems.

The next frontier of time-varying photonics lies not only in achievingfaster and stronger modulation, but also in understanding how driven materials interact with electromagnetic fields, and what new phenomena emerge from this interplay, may establish time-varying photonics as a broader platform for exploring nonequilibrium electrodynamics.
We hope this review has conveyed not only the common physical principles of these developments, but also the opportunities opened by treating time as a genuine dimension in design. 
With future advances in materials, modulation strategies, and programmable architectures \cite{wang2025reconfigurable, ma2026reconfigurable}, we anticipate that time-varying photonics will become a powerful framework for exploring fundamental physics and reshaping photonic technologies.

\subsection*{Disclosures}
The authors declare that there are no financial interests, commercial affiliations, or other potential conflicts of interest that could have influenced the objectivity of this research or the writing of this paper.

\subsection* {Code, Data, and Materials Availability} 
This review article did not generate any original data or code.
All data discussed are available in the cited publications.

\subsection* {Acknowledgments}
The research was supported by National Key R\&D Program of China (No. 2023YFA1407200) and National Natural Science Foundation of China (12192252).


\bibliography{ref}   
\bibliographystyle{spiejour}   

\end{spacing}
\end{document}